%% file: InfPC.tex
\documentclass[prd,twocolumn,amsmath,amssymb,floatfix,superscriptaddress,nofootinbib,preprintnumbers]{revtex4-1}
 \usepackage[utf8]{inputenc}
\usepackage{graphicx}
\usepackage{amssymb}
\usepackage{amsmath}
\usepackage{bm}
\usepackage{color}
\usepackage{enumitem}

\DeclareMathAlphabet\mathbfcal{OMS}{cmsy}{b}{n}
\definecolor{darkgreen}{RGB}{50,150,0}
\definecolor{purple}{cmyk}{0.5,0.75,0,0}
\definecolor{darkpurple}{RGB}{128,0,128}

\newcommand{\curv}{{\cal R}}
\newcommand{\Planck}{{\it Planck }}
\newcommand{\Hunits}{km\,s$^{-1}$Mpc$^{-1}$ }
\newcommand{\Comment}[1]{{}}

\definecolor{ultramarine}{rgb}{0.07, 0.04, 0.56}
\definecolor{cadmiumgreen}{rgb}{0.0, 0.42, 0.24}
\definecolor{indigo(dye)}{rgb}{0.0, 0.25, 0.42}
\usepackage[linktocpage=true]{hyperref}
\hypersetup{
colorlinks=true,
citecolor=ultramarine,
linkcolor=cadmiumgreen,
urlcolor=indigo(dye),
pdfauthor={},
pdftitle={},
pdfsubject={}
}

\begin{document}
\preprint{YITP-SB-16-18}

\title{Inflationary Features and Shifts in Cosmological Parameters from \Planck 2015 Data}

\author{Georges Obied}

\affiliation{Harvard University, Department of Physics, \\Cambridge, MA 02138, USA}

\author{Cora Dvorkin}

\affiliation{Harvard University, Department of Physics, \\Cambridge, MA 02138, USA}

\author{Chen Heinrich}

\affiliation{Kavli Institute for Cosmological Physics, Department of Astronomy \& Astrophysics,  Enrico Fermi Institute, University of Chicago, Chicago, IL 60637}

\author{Wayne Hu}

\affiliation{Kavli Institute for Cosmological Physics, Department of Astronomy \& Astrophysics,  Enrico Fermi Institute, University of Chicago, Chicago, IL 60637}

\author{Vinicius Miranda}

\affiliation{Center for Particle Cosmology, Department of Physics and Astronomy,\\University of Pennsylvania, Philadelphia, Pennsylvania 19104, USA}

\begin{abstract}
We  explore the relationship between features in the \Planck 2015 temperature and polarization data,  shifts in the cosmological
parameters, and  features from inflation.  Residuals in the temperature data at low multipole $\ell$,
which are responsible for the high $H_0\approx 70$ \Hunits and low $\sigma_8\Omega_m^{1/2}$ values from $\ell<1000$ in power-law $\Lambda$CDM models,
are better fit to inflationary features with a $1.9\sigma$ preference for running of the running of the tilt or a stronger
$99\%$ CL local significance preference for a sharp drop in power around $k=0.004$\,Mpc$^{-1}$ in generalized slow roll
and a lower $H_0\approx 67$ \Hunits.   The same in-phase acoustic residuals at $\ell>1000$ that drive the global $H_0$ constraints and appear as a lensing anomaly also favor running parameters which allow even lower  $H_0$, but not once lensing reconstruction is considered.  Polarization spectra are intrinsically highly sensitive to these parameter shifts, and even more so in the \Planck 2015 TE data due to an outlier at $\ell \approx 165$, which disfavors the best fit $H_0$ $\Lambda$CDM solution by more than $2\sigma$,
and high $H_0$ value at almost $3\sigma$.  Current polarization data also slightly enhance the significance of a sharp
suppression of large-scale power but leave room for large improvements in the future with cosmic variance limited $E$-mode measurements.
\end{abstract}

\maketitle

\section{Introduction}

The anisotropies of the cosmic microwave background (CMB) continue to be one of the most powerful probes we have to study physical conditions in the early universe. With the release of {\it Planck} 2015 data, we now have access to precise all-sky measurements of the polarization as well as the temperature fluctuations of the CMB. Observations from supernovae, baryon acoustic oscillations, Big Bang Nucleosynthesis, among other datasets support, in addition to the CMB, a broadly consistent cosmological model termed $\Lambda$CDM.

Nonetheless, evidence for features beyond $\Lambda$CDM  in the CMB temperature data have been claimed
(e.g.~\cite{Bennett:1996ce,Spergel:2003cb,Hinshaw:2003ex,Peiris:2003ff,Mortonson:2009qv,Ade:2013kta,Ade:2013zuv,Cai:2015xla,Ade:2015xua}), as well as disagreements between  CMB predictions for local observables under $\Lambda$CDM and their measured
values (e.g.~the local expansion rate $H_0$ \cite{Ade:2013zuv,2011ApJ...730..119R}).
The two are related because these local cosmological inferences depend on the assumptions in the
$\Lambda$CDM model, in particular the form of the inflationary power spectrum.
It is, thus, essential to consider the impact of one type of deviation from $\Lambda$CDM on the other.   Moreover polarization data play an essential
role in breaking degeneracies between the two.
It is the aim of this paper to examine these relationships between features in the temperature and polarization data, shifts in cosmological parameters and inflationary features.

Previous studies have tested the consistency of {\it Planck} data at different stages in the chain of inference from the raw data to cosmological parameters  (e.g.~\cite{Larson:2014roa,Spergel:2013rxa,addison2016}).
In particular,
Ref.\ \cite{addison2016} split the temperature power spectrum into two disjoint multipole ranges, the lower being similar to the range of WMAP, and analyzed the two ranges separately. They then find a $\sim 2-3\sigma$  discrepancy between parameters derived from the two parts of the same dataset. The \Planck collaboration \cite{aghanim2016} carried out a meticulous investigation of the cause of this tension and discovered that it was mainly due to a deficit in power at low-$\ell$ and oscillatory residuals in the multipole range $\ell\gtrsim 10^3$. More recently, Ref.~\cite{Shafieloo:2016zga} examined the consistency of the polarization and temperature datasets, finding no strong evidence
to disfavor the $\Lambda$CDM model.

In addition, several works \cite{Hu:2003vp,Bridle:2003sa,Leach:2005av,Sealfon:2005em,Dvorkin:2007jp,Mortonson:2009qv,2010PhRvD..81b1302P,2010ApJ...711....1P,Dvorkin:2011ui,Planck:2013jfk,miranda2015,Ade:2015lrj}
 have explored the effects of inflationary features on CMB observables, including the central role polarization data play in confirming or rejecting such features.  For instance, step-like features in the inflationary potential could cause a dip and bump in the temperature data at low $\ell$.
Such models could arise from symmetry-breaking phase transitions in the early universe, among other reasons \cite{Silk:1986vc,Polarski:1992dq,Adams:1997de,Hunt:2004vt}. These features invariably violate the slow-roll approximation and, thus, require more sophisticated modeling of the resultant inflationary power spectrum.

 In this paper, we adopt both the traditional running of the tilt type parameters
and the generalized slow roll (GSR) formalism \cite{Stewart:2001cd,Dvorkin:2009ne,Dvorkin:2010dn,Dvorkin:2011ui}, which allows the inclusion of extra `spline basis' parameters that accommodate order unity power spectrum variations from slow roll predictions.  We study the shifts in cosmological
parameters, such as $H_0$ and the amplitude of local structure, when these extra parameters are included and explore the aspects of the temperature and polarization data that drive them.

This paper is organized as follows. Sec.\ \ref{sec:data} describes the datasets and parameter inference techniques. The results of our analyses are presented in Sec.\ \ref{sec:H0}, which focuses on the shifts in cosmological parameters, and in Sec.\ \ref{sec:inflation}, which focuses on the implications for inflationary features. We conclude in Sec.\ \ref{sec:discussion}.

\begin{table}
\begin{tabular}{|l|l|}
  \hline
  Dataset & Likelihood \\
  \hline \hline
  TT & binned TT + low TEB \\
  TTEE & binned TTTEEE + low TEB \\
  $\phi\phi$ &  lens reconstruction\\
    \hline
\end{tabular}
\caption{\footnotesize Combinations of  {\it Planck} 2015 data sets and
likelihoods.\footnote{Specifically:\\
low TEB\,$=\texttt{lowl\_SMW\_70\_dx11d\_2014\_10\_03\_v5c\_Ap}$\\
binned TT\,$=\texttt{plik\_dx11dr2\_HM\_v18\_TT}$\\
binned TTTEEE\,$=\texttt{plik\_dx11dr2\_HM\_v18\_TTTEEE}$\\
lens reconstruction\,$=\texttt{smica.g30\_ftl\_full\_pp}$, $40\le \ell \le 400$}  In addition, we split the binned TT and TTTEEE likelihoods multipole ranges,
the latter with a modified likelihood code.  }
  \label{tab:data}
\end{table}

\begin{table}
\begin{tabular}{|l|l|}
  \hline
  Model & Parameters $\{\tau, \theta_{\rm MC}, \Omega_bh^2, \Omega_ch^2 \}$\\
  \hline \hline
  $\Lambda$CDM & $+\{A_s, n_s\}$ \\
  r$\Lambda$CDM & $+\{A_s, n_s, n_{\rm run}, n_{\rm run,run}\}$\\
  SB & $+\{A_s, n_s, p_i\}$ \\
  rSB & +$\{A_s, n_s, n_{\rm run}, n_{\rm run,run}, p_i\}$ \\
  \hline
\end{tabular}
  \caption{\footnotesize Models and parameterizations of the inflationary power spectrum.
  The parameters $p_i$ are spline basis coefficients that generalize the tilt as defined in Appendix~\ref{app:GSR}.
  In each case, the 4 late time cosmological parameters are prepended to the parameter list.
 }
   \label{tab:models}
\end{table}

\begin{figure*}
\centering
\includegraphics[width=\textwidth]{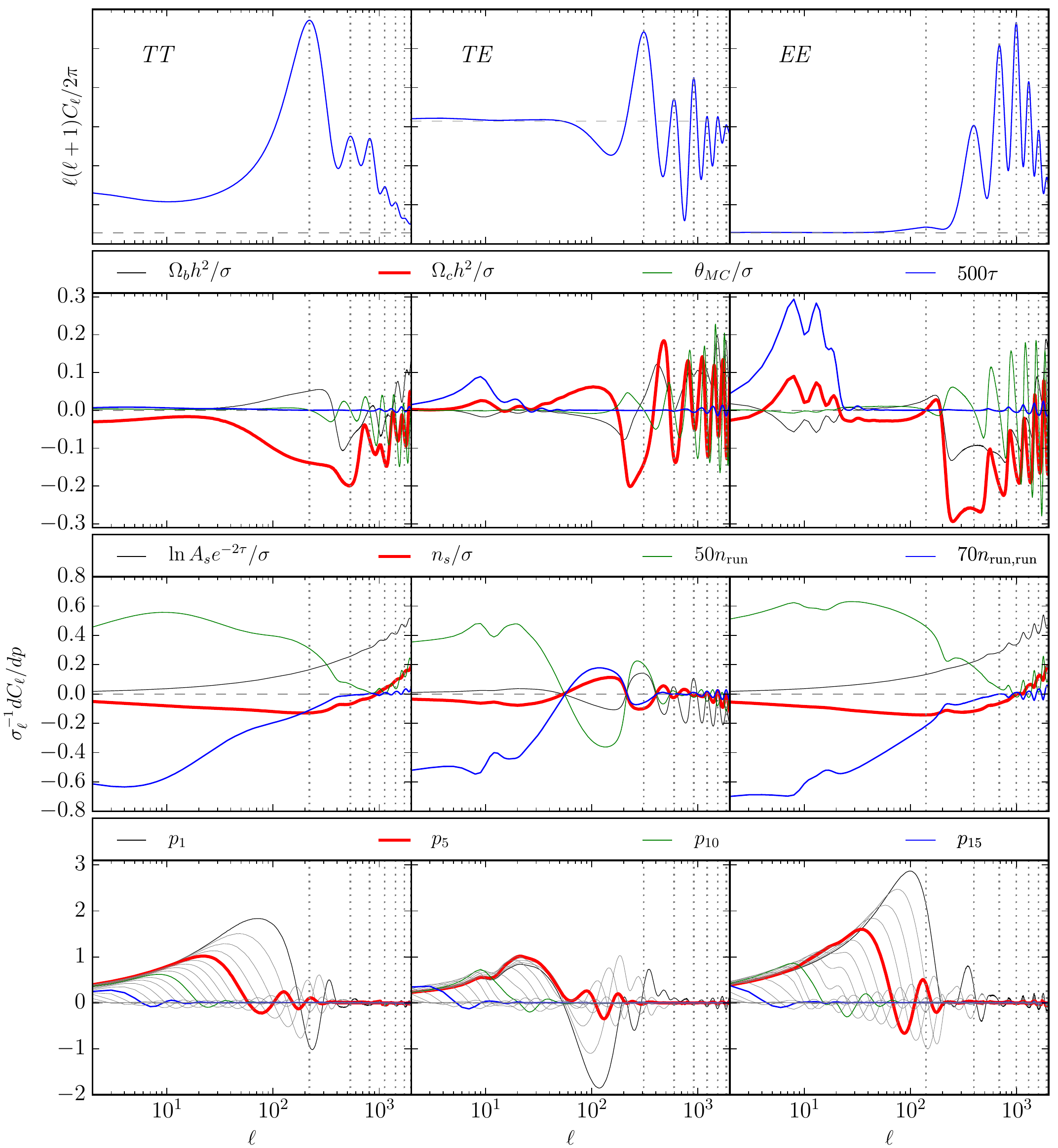}
\caption{\footnotesize Response of the TT, TE, EE power spectra to various parameters around the $\Lambda$CDM TT model of Tab.~\ref{tab:modelstats}.  Derivatives are scaled by $\sigma_\ell$, the cosmic variance
error per $\ell$ of the model, and parameters are scaled by either their \Planck 2015 $\Lambda$CDM TT
error $\sigma$ or by a fixed
number for visibility.  Vertical dotted lines are drawn at the acoustic peak locations and
horizontal dashed lines at zero levels.}
\label{fig:response}
\end{figure*}

\section{Data and Models}
\label{sec:data}

For the analyses described in this paper, we use the publicly available {\it Planck} 2015 data, which include the power spectra of cosmic microwave background (CMB) temperature and polarization fluctuations. For the low multipole $(2\le \ell\le 29)$ range, we use the standard \texttt{plik lowTEB} likelihood code supplied by the {\it Planck} collaboration. For the high multipole $(\ell\ge 30)$ range, we similarly use the {\it Planck} \texttt{plik} binned\footnote{We have separately tested  that the unbinned likelihoods give statistically indistinguishable results.}   TT likelihood for the baseline and the TTTEEE likelihood to assess the impact of the polarization data.   Note that the low-$\ell$ polarization data are included in all analyses, even though we refer to these cases as ``TT" and ``TTEE" respectively.
We also alter the $\ell$ range of the different likelihoods to probe the impact of
specific regions of the data, in the case of the TTTEEE likelihood using a custom modification of the code.  In some cases, we use information from the  {\it Planck}  lensing power spectrum $C_\ell^{\phi\phi}$ in the multipole range $40 \le \ell \le 400$.  We also include the standard foreground parameters in
each analysis.\footnote{Premarginalizing foregrounds assuming $\Lambda$CDM with
\texttt{plik-lite} as in Ref. \cite{aghanim2016} can change inferences on the inflationary running parameters.   Conversely, not using the whole range in $\ell$ to constrain foregrounds allows
$\ell < 1000$ models that do not provide a good global fit to foregrounds (see Tab.~\ref{tab:modelstats} and \cite{aghanim2016}).}
 These datasets and likelihoods are summarized in Tab.~\ref{tab:data}.

We employ a Bayesian approach to infer parameter posterior distributions and to derive confidence limits,
namely the Markov Chain Monte Carlo (MCMC) technique implemented via the publicly available \texttt{CosmoMC} code~\cite{Lewis2002} linked to a modified version of the Boltzmann code \texttt{CAMB}~\cite{Lewis1999}. For each combination of datasets and models, we run 4 MCMC chains until convergence, determined by the Gelman-Rubin criterion $R-1<0.01$.

For the parameters, we take in all cases the standard 4 late-time
$\Lambda$CDM cosmological parameters: the optical depth to
reionization $\tau$, the (approximate) angular scale of the sound horizon $\theta_{\rm MC}$,
 the baryon density $\Omega_b h^2$ and the cold dark matter density $\Omega_c h^2$.
 Tensions in the $\Lambda$CDM cosmology are mainly associated with $\Omega_c h^2$, which
 determines the calibration of the physical scale of the sound horizon and, hence, the CMB inference of
 the Hubble constant $H_0$ as well as the growth of structure and lensing observables
 represented by $\sigma_8 \Omega_m^{1/2}$.   These two parameters expose tensions in the $\Lambda$CDM model when compared to local,
  non-CMB measurements.   Here and throughout the paper, $\sigma_8$ is the rms of the linear density field smoothed with a tophat at  $8 h^{-1}$ Mpc and $\Omega_m = \Omega_c+\Omega_b+\Omega_\nu$.   We assume one massive neutrino with $m_\nu = 0.06$ eV
  and the usual  $N_{\rm eff} = 3.046$.  We summarize our model and parameter choices in Tab.~\ref{tab:models}.

\begin{table*}
  \begin{tabular}{|c|c|c|c|c|c|c|c|c|}
    \hline
    \input{table1_3.tex}
  \end{tabular}
  \caption{Cosmological parameters of best fit models for various datasets. In the $\Lambda$CDM TT $(\ell < 1000)$ model the best fit assumes the foreground model of the best fit $\Lambda$CDM TT model,
  unlike the MCMC results where foregrounds are jointly fit.  This assures a good global fit when foregrounds are ill-constrained by subsets of the data.
}
   \label{tab:modelstats}
\end{table*}

 In Fig.~\ref{fig:response} we show the parameter
  derivatives or responses of the TT, TE and EE power
 spectra around the best fit $\Lambda$CDM model to the TT \Planck dataset given in
 Tab.~\ref{tab:modelstats}.    Here
 we normalize the derivatives to $\sigma_{\ell}$, the cosmic variance per multipole, where
 \begin{equation}
 \sigma_{\ell} =
 \begin{cases}
\sqrt{\frac{2}{2\ell+1}} C_\ell^{TT}, & TT;\\
\sqrt{\frac{1}{2\ell+1}}  \sqrt{ C_\ell^{TT} C_\ell^{EE} + (C_\ell^{TE})^2}, & TE;\\
\sqrt{\frac{2}{2\ell+1}}  C_\ell^{EE}, & EE. \\
\end{cases}
\end{equation}
Here and below $\sigma_\ell$ is fixed by the fiducial TT model when varying parameters.
  For TT and EE this rescaling
produces fractional derivatives scaled by $\sim \ell^{-1/2}$ at $\ell\gg 1$, which has the benefit of
removing features from parameters like normalization that simply echo the acoustic peaks in the spectrum.
We have also rescaled the parameters to either the error
in the parameter from the $\Lambda$CDM TT model, e.g.\ for $\Omega_ ch^2$ we show the parameter $p=\Omega_c h^2/
\sigma(\Omega_c h^2)$, or by a fixed number, e.g.\ for $n_{\rm run}$, $p=50 n_{\rm run}$ for clarity.
For the cases where we scale to $\sigma$, this has the added benefit of showing where
current limits can be improved with better measurements.

Increasing $\Omega_c h^2$ decreases the overall power in the first few acoustic
 peaks in TT due to the reduction of radiation driving and the early ISW effects from decaying potentials inside
 the sound horizon (see \cite{Hu:1994uz} Fig.~4).  Since the former effect changes the amplitude of acoustic oscillations, it also reduces
 the temperature peaks, which are extrema of the oscillations, relative to troughs, which are nodes.
 Before the damping tail, these changes are nearly in phase with the acoustic peaks  in contrast to $\theta_{\rm MC}$, which makes out-of-phase changes that shift peak positions.  Damping
 tends to shift the observed peak locations to larger angles, the  phasing of the two drifts
 at high $\ell$, but we will nonetheless refer to them as ``in phase" and ``out of phase" oscillatory
 changes since our focus is on the peaks that are well measured by Planck.   Note that with
 higher multipole information from ground based experiments this effect can be distinguished
 from entirely in phase effects.
 Indeed increasing $\Omega_c h^2$ also increases the gravitational lensing, which also reduces the peaks relative to the troughs and remains entirely in phase through the damping tail.
  For the \Planck data, the acoustic effect of $\Omega_c h^2$ is larger than the lensing effect but they both contribute \cite{aghanim2016}.

  The impact of $\Omega_c h^2$ on the EE spectrum is much sharper near the sound horizon ($\ell\approx 200$, first polarization trough, see Fig. 3 in Ref. \cite{Hu:2001bc}) since potential
 decay does not change it after recombination as it does temperature through the early ISW effect.
 This enhanced sensitivity carries over to the TE spectrum, which dominates the \Planck polarization
 constraints at high $\ell$.

To infer the 4 late-time parameters from CMB data we must assume a parameterized form for the inflationary
curvature spectrum $\Delta_{\cal R}^2$, and vice versa.  We choose to model $\Delta_{\cal R}^2$
 in 4 ways to highlight its impact.
In all cases, $A_s$ is the amplitude $\Delta_{\cal R}^2(k_0)=A_s$ where
 $k_0=0.08$~Mpc$^{-1}$.   Fig.~\ref{fig:response} shows that this choice places the pivot point at $\ell \approx 10^3$, which is near the best constrained scale of the \Planck data.
It also shows the effect of raising lensing by raising $A_s$  in the oscillatory response beyond the third peak.

For the slope $d\ln \Delta_{\cal R}^2/d\ln k$, we consider a
constant tilt $n_s-1$ as the baseline case, which we call ``$\Lambda$CDM".
Unlike the late-time cosmological parameters, $A_s$ and $n_s$ coherently change the TT  spectrum across all multipoles rather than introducing localized  features in
$\Delta C_\ell^{\rm TT}/\sigma_\ell$ (see Fig.~\ref{fig:response}).  On the other hand, the dominant source of tension in the
$\Lambda$CDM model comes from anomalies in the TT spectrum that are more localized.
Fig.~\ref{fig:response} shows that by allowing running and ``running of the running" of the tilt:

\begin{equation}
\frac{d \ln \Delta_{\cal R}^2}{d\ln k} = n_s -1 + n_{\rm run}\ln\left(\frac{k}{k_0}\right)
 + \frac{n_{\rm run,run}}{2}\ln^2\left(\frac{k}{k_0} \right),
 \label{eqn:running}
\end{equation}
one can change the low-multipole part of the spectrum relative to the high-multipole part, but only if these additional
parameters are of order the $n_s-1$ itself.
 In standard slow roll inflation, each order of running is suppressed by an additional
 factor of $n_s-1$, which should make them safely unobservable in the \Planck data.
 We call the  case where these parameters are not constrained by the slow roll hierarchy ``r$\Lambda$CDM".  In these models, $\Delta_\curv^2$ will have a prominent but broad feature on
 CMB scales.

\begin{figure*}
\centering
\includegraphics[width=0.85\textwidth]{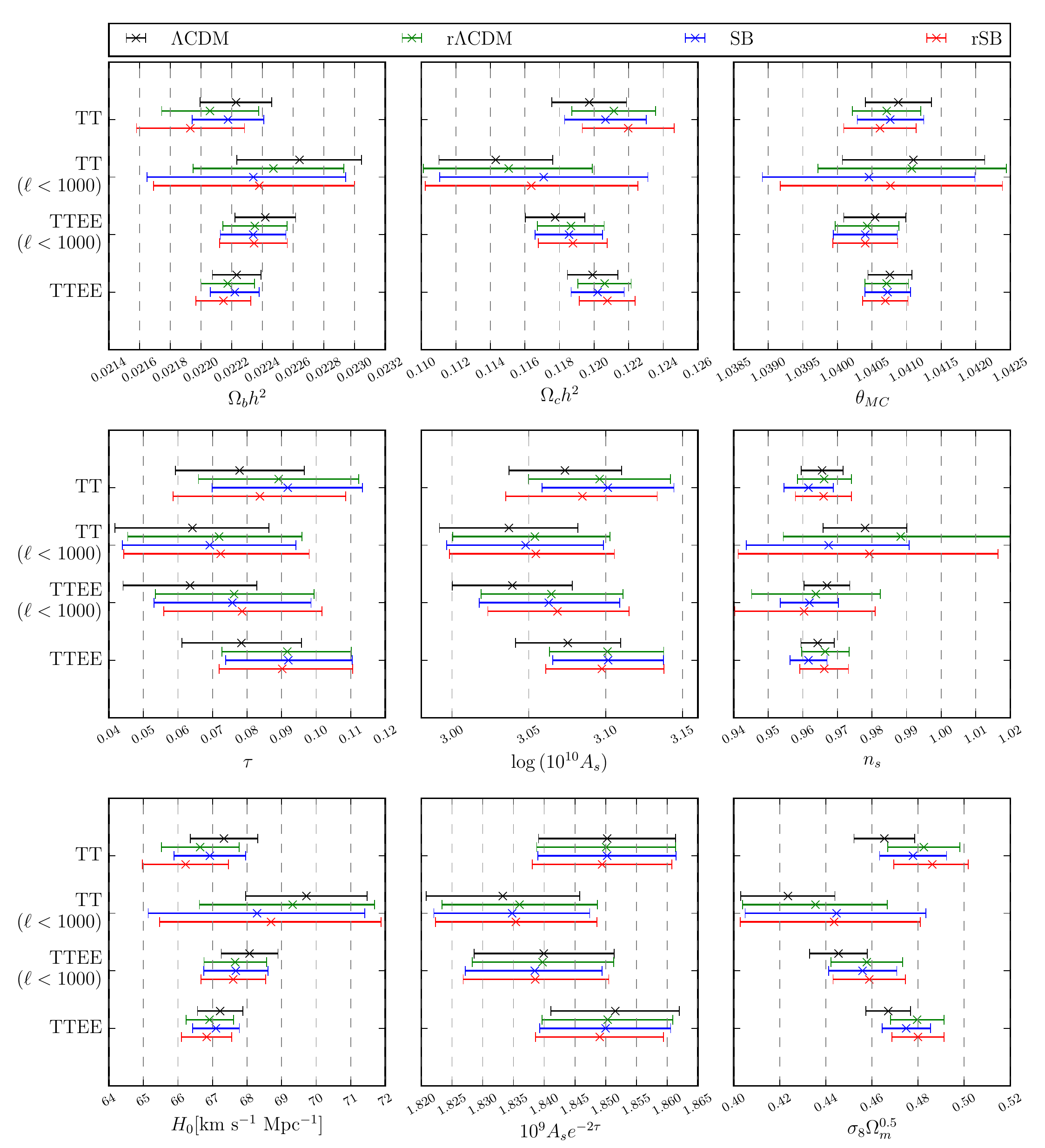}
\caption{\footnotesize Cosmological parameter constraints with the 4 combinations of datasets
in Tab.~\ref{tab:data} on the 4  models for the inflationary power spectrum in Tab.~\ref{tab:models}. }
\label{fig:boxplot}
\end{figure*}

\begin{figure*}
\centering
\includegraphics[width=\textwidth]{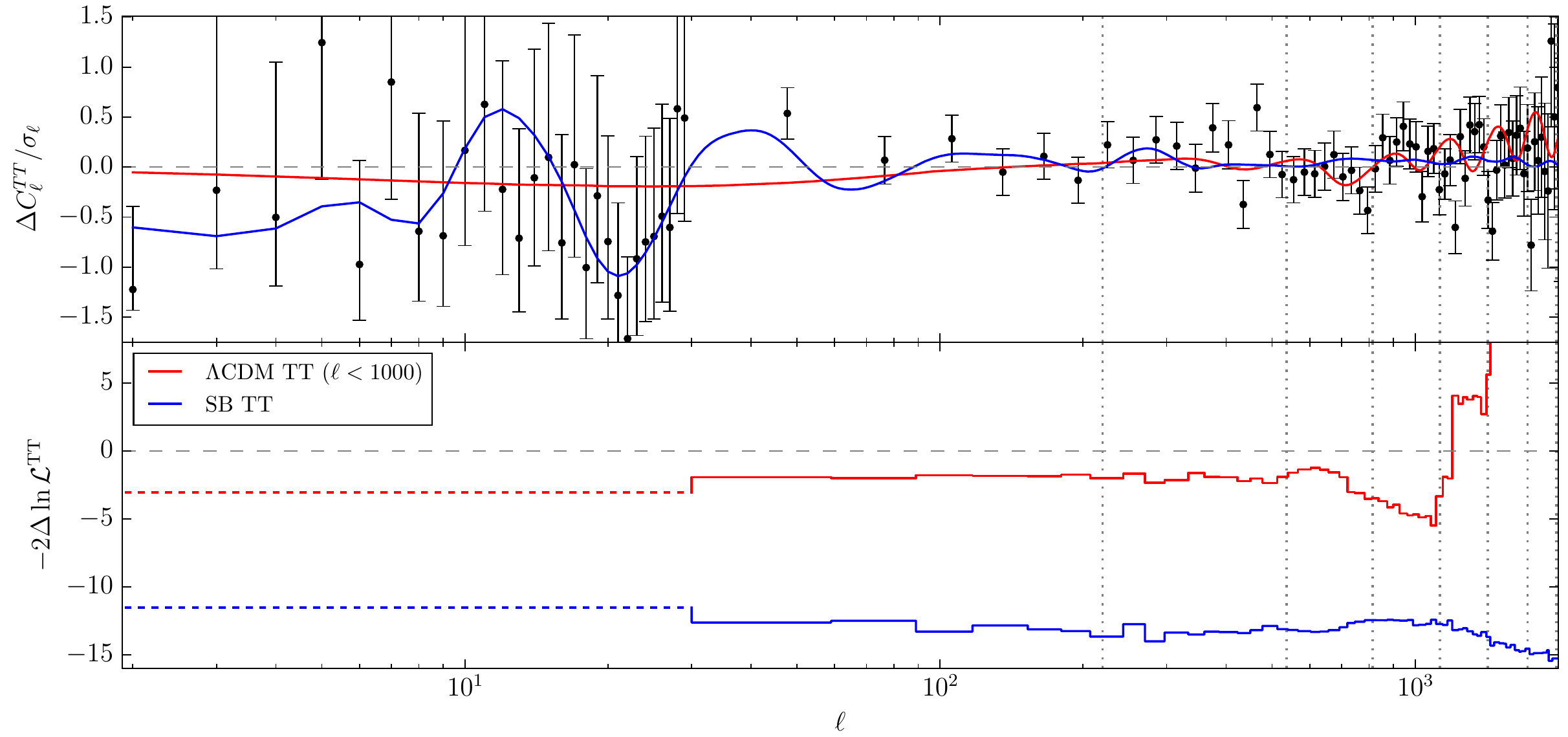}
\caption{\footnotesize Top panel: residuals in the TT data (points) and models relative to the best fit $\Lambda$CDM TT model of Tab.~\ref{tab:modelstats} (dashed line).  Bottom panel: change in the  likelihood $2\Delta \ln {\cal L}^{\rm TT}$
 of models relative to
$\Lambda$CDM TT starting from the lowest bin at $2\le \ell<30$ (dotted lines) and adding each  individual data point thereafter.
The $\Lambda$CDM TT ($\ell < 1000$) model (red line) fits  the  residuals at low $\ell \lesssim 40$ and $\ell \sim 700-1000$ better due to changes in $H_0$ and other parameters, but it is a poor fit to the in-phase residuals at $\ell > 1000$.   Conversely, the inflationary feature
SB model (blue) fits the low-$\ell$
residuals yielding a large $2\Delta \ln {\cal L}^{\rm TT} \sim 12$ improvement at low $\ell$, but
only small changes at high $\ell$.  Note that the $\Lambda$CDM TT ($\ell<1000$) model has the same foreground parameters as the $\Lambda$CDM TT model, and the data points assume TT foregrounds are fixed to the best fit $\Lambda$CDM TT values in Ref. \cite{Ade:2015xua}.}
\label{fig:TTresiduals}
\end{figure*}

While these parameters can model large-amplitude deviations from slow roll, they cannot model
rapid variations that occur on a time scale of an efold or shorter.  Such rapid variations would be required to
produce  sharp features in multipole space in the
observed power spectra.  To distinguish between rapid variations that are confined to
low multipoles and a smooth change in the overall amount of large to small-scale power, we
employ the
 generalized slow-roll (GSR) approximation.   Here,
\begin{equation}
\frac{d \ln \Delta_{\cal R}^2}{d\ln k}  \rightarrow - G',
\end{equation}
where $G'$ is a function of efolds of the sound horizon during inflation, as detailed in Appendix~\ref{app:GSR}.   We consider
arbitrary variations $\delta G'$ around both the standard $n_s-1$ (``SB") and running forms (``rSB") for
scales between the horizon and the first acoustic peak.
Specifically, we take a spline basis with 20 knots $p_i$ across 2 decades in scale.
In Fig.~\ref{fig:response} we show the response of the observable power spectrum to these parameters.  Note that individually these parameters produce step-like features in the power spectrum
and their superposition can produce fine scale features to fit the low-$\ell$ TT anomalies in the data.
In Fig.~\ref{fig:response} these steps appear tapered due to rescaling by cosmic variance
errors $\sigma_\ell^{-1} \propto \ell^{1/2}$.
They produce sharper features in EE at a slightly higher multipole due to projection effects.
Since the GSR approximation is still perturbative, we impose an integral constraint on the second order
impact of the
$p_i$ parameters,  $\mathrm{max}(I_1) < {1}/{\sqrt{2}}$ (see Eq.~\ref{eq:I1prior}).
Finally, we assume no tensors, $r=0$ throughout.

Cosmological parameter constraints derived from the MCMC likelihood analyses are summarized in Fig.~\ref{fig:boxplot}, and best fit models are listed in Tab.~\ref{tab:modelstats}.
We use these models in the next two sections to highlight the interplay of features in the
data, shifts in cosmological parameters, and features in inflation.

\begin{figure}
\centering
\includegraphics[width=0.4\textwidth]{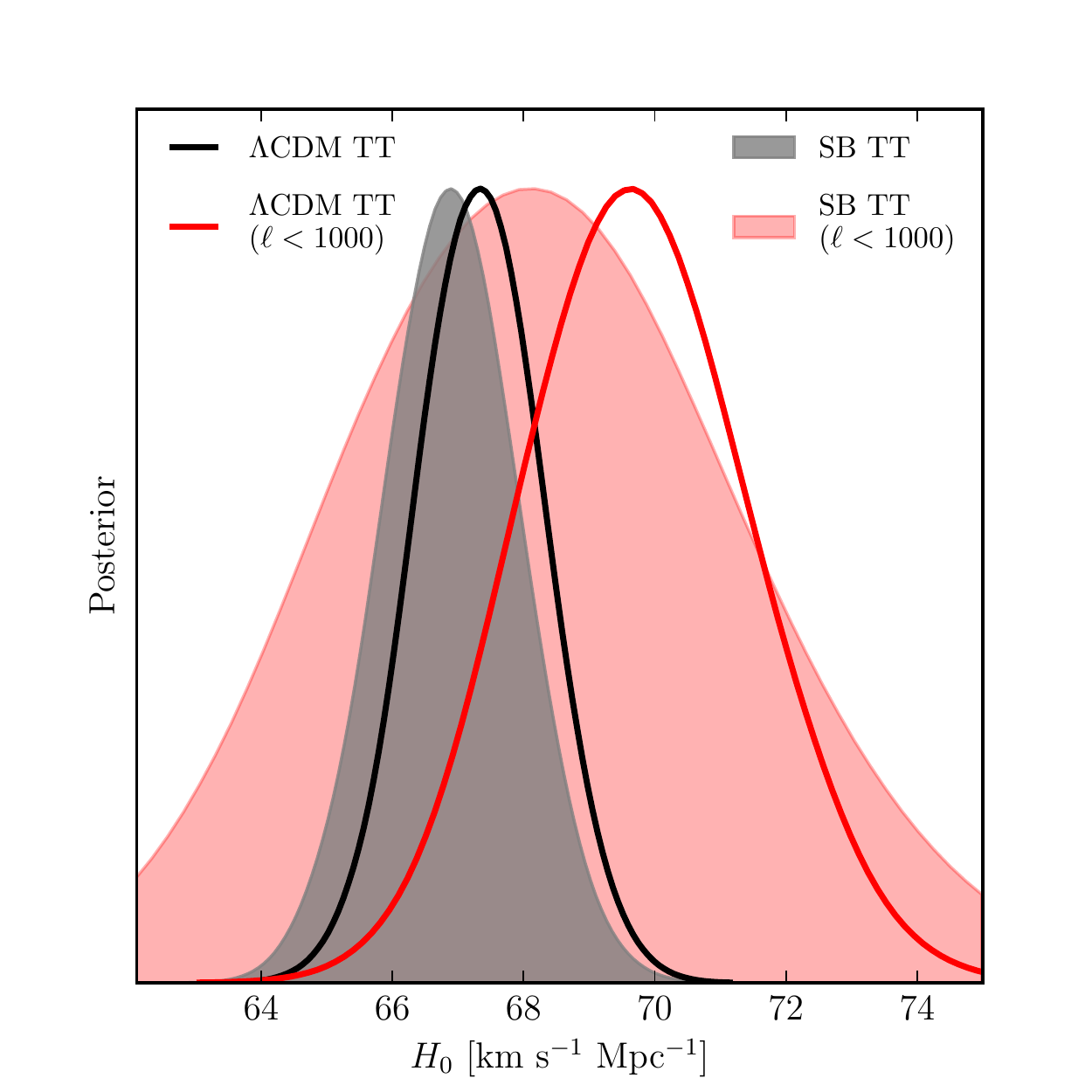}
\caption{\footnotesize $H_0$  from the TT  full vs $(\ell <1000)$ datasets.
Low-$\ell$ residuals in Fig.~\ref{fig:TTresiduals} drive $H_0$ higher in the $\Lambda$CDM TT
$(\ell <1000)$ analysis (red vs black curves). In the SB model, these residuals are fit by inflationary rather than cosmological parameter changes, so low values of $H_0$ that are favored in the
full analysis (gray shaded) are also compatible in the
$(\ell<1000$, red shaded) analysis.  }
\label{fig:H0_lowl}
\end{figure}

\begin{figure}
\centering
\includegraphics[width=0.49\textwidth]{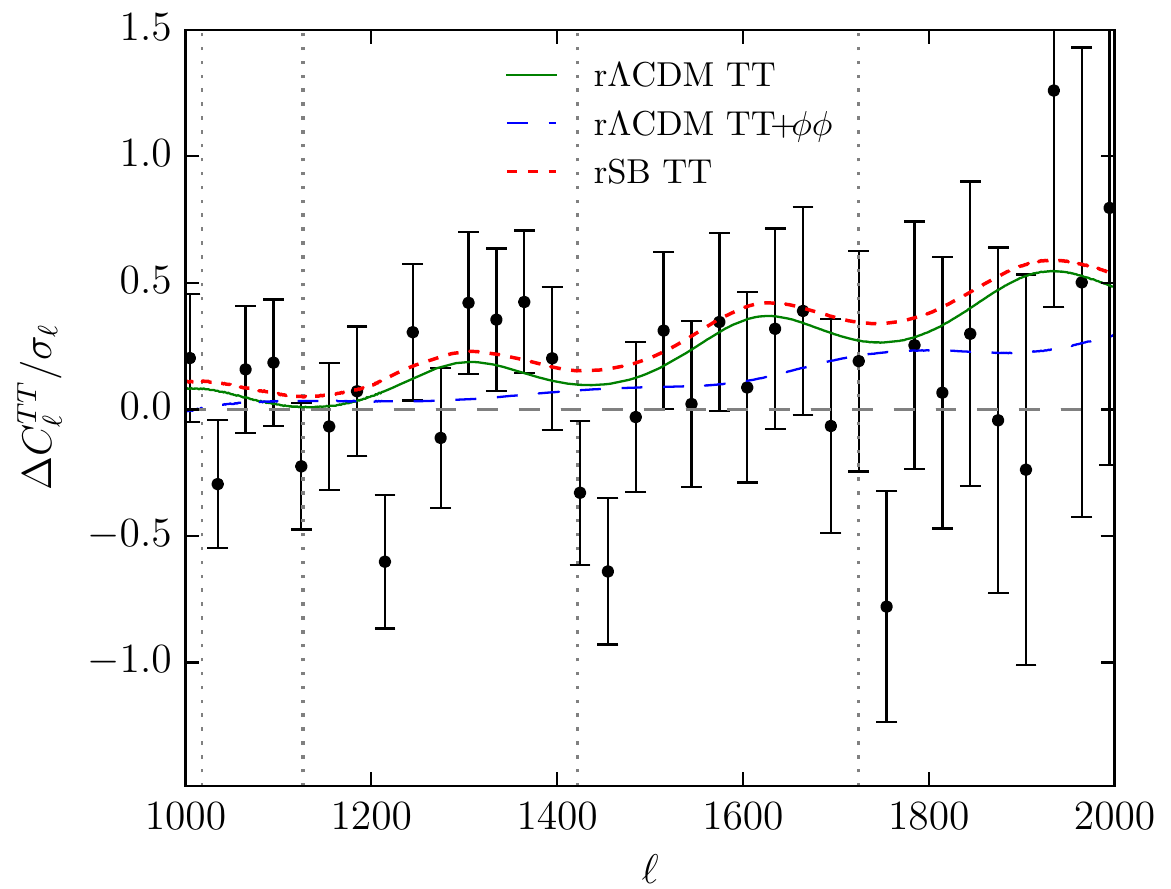}
\caption{\footnotesize TT residuals at $\ell>1000$ relative to the best fit $\Lambda$CDM TT model of Tab.~\ref{tab:modelstats} (dashed line).   With the running parameters, r$\Lambda$CDM
and rSB seek to fit the oscillatory residuals and make the peaks smoother.   Adding the
lens reconstruction data $\phi\phi$ largely removes this preference.
}
\label{fig:highLresid}
\end{figure}

\section{$H_0$ and Cosmological Parameters}
\label{sec:H0}

Of the 4 late-time parameters, the one most dependent on the
dataset and model assumptions of the previous section
 is the cold dark matter density $\Omega_c h^2$ (see
 Fig.~\ref{fig:boxplot}), which directly controls the inferences on $H_0$ and
the amount of structure today $\sigma_8 \Omega_m^{1/2}$.   Raising $\Omega_c h^2$
increases the scale of the sound horizon {\it relative} to the current horizon because of the
reduced effect of radiation on the former.  Compensating this effect  requires a lower Hubble constant 
 to increase the distance to recombination. In
$\Lambda$CDM the net effect is to keep $\Omega_m h^3$ nearly constant \cite{Hu:2000ti}.

As discussed in Ref.~\cite{aghanim2016}, this sensitivity is largely driven by the \Planck TT  residuals from the  $\Lambda$CDM best fit model in two regions: power deficits and glitches at
low $\ell\lesssim 40$ and oscillatory residuals at $\ell \gtrsim 10^3$.
As shown in Fig.~\ref{fig:response}, $\Omega_c h^2$ controls the overall amplitude of the acoustic peaks relative to low multipoles, and also impacts the height of the peaks relative to the troughs.      Given the ability of the inflationary power spectrum to change the relative power and amount of smoothing of the peaks due to lensing,
the interpretation of these residuals
depends on the inflationary model assumptions.     In addition, as shown in Ref.~\cite{Ade:2015xua}, the TE data have a surprisingly large impact on $\Omega_c h^2$ and $H_0$,
favoring a low value for the latter.    We further explore these issues and their relationship to inflationary features in turn.

\begin{figure}
\centering
\includegraphics[width=0.4\textwidth]{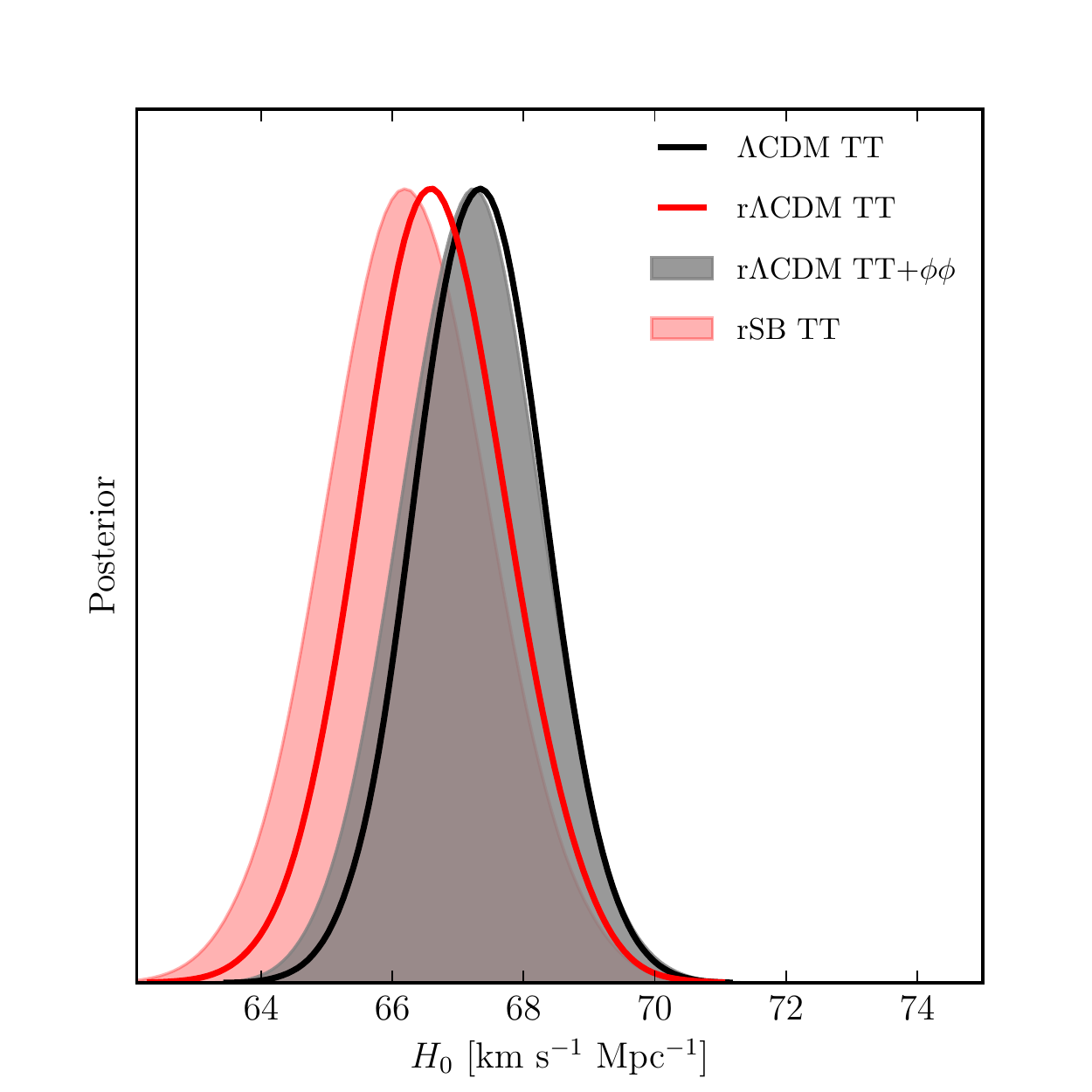}
\caption{\footnotesize $H_0$ and inflationary parameters.  In phase $\ell>1000$ residuals in Fig.~\ref{fig:TTresiduals} drive $H_0$ lower than the $\Lambda$CDM TT model (black) but require running parameters
 (red curve, r$\Lambda$CDM; red shaded, rSB)
in order to produce a global fit.  These fits also predict more gravitational lensing and are not favored by lensing reconstruction data (gray shaded, r$\Lambda$CDM TT+$\phi\phi$.)  }
\label{fig:H0_lens}
\end{figure}

\subsection{Low $\ell \lesssim 40$ TT residuals}

We begin by repeating the test discussed in Ref.~\cite{addison2016} of comparing the inferences from the TT and  TT ($\ell<1000$)
  datasets on $H_0$ in the $\Lambda$CDM model context.
In Fig.~\ref{fig:H0_lowl}, we show that constraints shift from $H_0=[67.3\pm0.98]$ \Hunits  to
$[69.7\pm 1.8]$ \Hunits, as expected.   In Ref.~\cite{aghanim2016}, this shift is largely
attributed to  the low-$\ell$ residuals
between the data and the best fit $\Lambda$CDM model to the full-TT data, which we display in
Fig.~\ref{fig:TTresiduals}.\footnote{These data points have a fixed best fit TT foreground model under $\Lambda$CDM subtracted.   In order to compare the $\Lambda$CDM TT ($\ell<1000$) model to the residuals, we take the best fit with the TT foreground model fixed by the full $\ell$ range in
Tab.~\ref{tab:modelstats}.}
These residuals favor lower low-$\ell$ power compared with the acoustic peaks.  As
seen in Fig.~\ref{fig:TTresiduals} (red curve), they can be fit better by the $\Lambda$CDM TT ($\ell<1000$) model by lowering
$\Omega_c h^2$, which increases the radiation driving to raise the peaks over lower $\ell$'s with adjustments of other parameters, such as $n_s$ (see Tab.~\ref{tab:modelstats}).
In addition, the out-of-phase residuals peaking between the third and fourth acoustic peaks
 $\ell \approx 700-1100$  drive a small change in $\theta_{\rm MC}$.
With the full range in TT, these same changes cannot fit the data, as we discuss further in the next
section.

\begin{figure}
\centering
\includegraphics[width=0.5\textwidth]{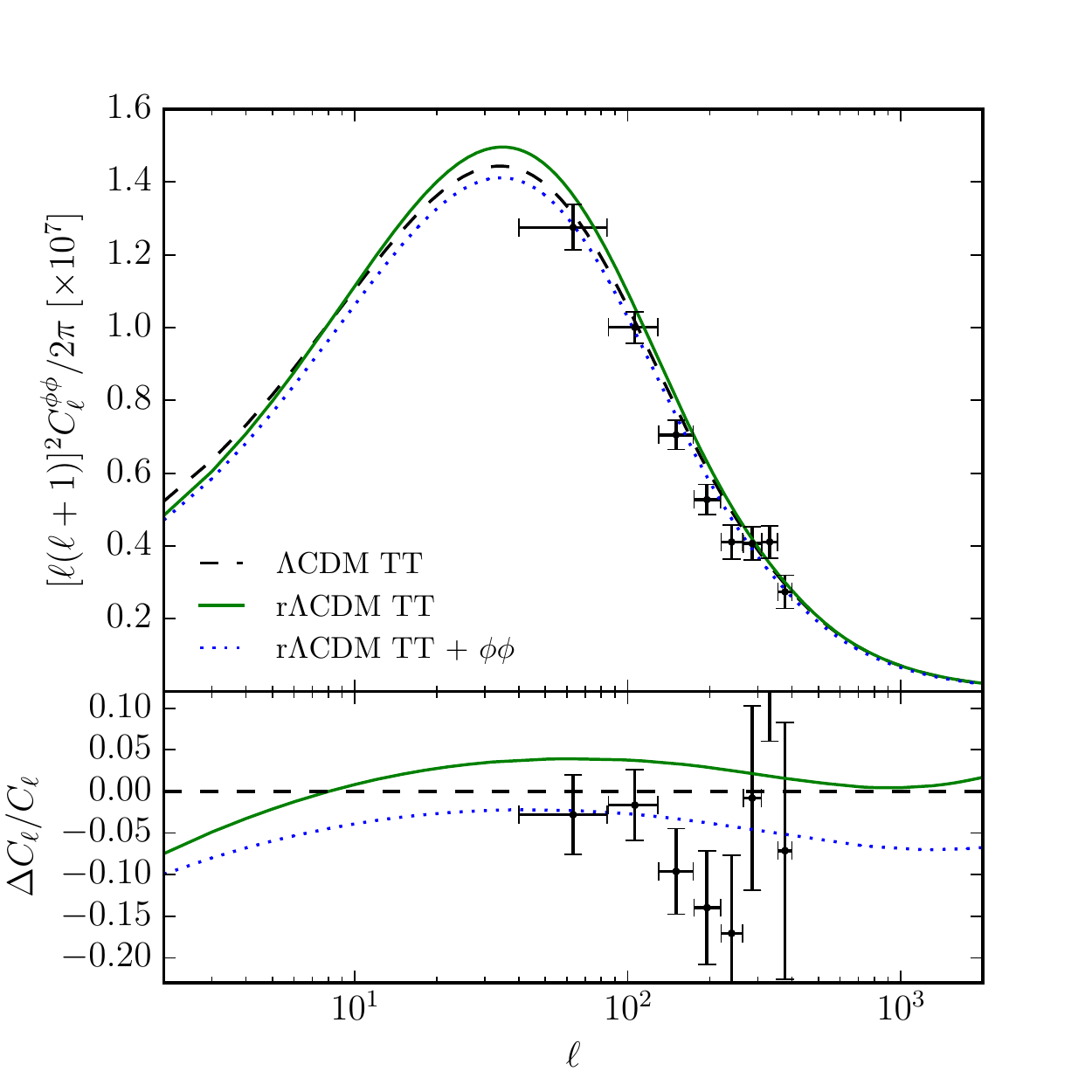}
\caption{\footnotesize Lensing power spectrum in the $\Lambda$CDM TT model vs. running models
r$\Lambda$CDM TT and r$\Lambda$CDM TT+$\phi\phi$.  Lowering $H_0$ in the $r\Lambda$CDM model simultaneously raises lensing and is not favored once the lensing reconstruction data (points)
are taken into account. }
\label{fig:ClPP}
\end{figure}

To further test this interpretation, we show the impact of generalizing the model class to SB which allows the low-$\ell$ residuals to be fit by inflationary features instead.
The maximum likelihood SB TT model of Tab.~\ref{tab:modelstats} is shown in Fig.~\ref{fig:TTresiduals}
(blue curve).   By design, the SB parameters do not significantly alter the acoustic peaks (see Fig.~\ref{fig:response}) leaving the high-$\ell$ residuals and cosmological parameters nearly
unchanged from $\Lambda$CDM TT.
 Correspondingly, in Fig.~\ref{fig:H0_lowl}
we show that for the full-TT data set, $H_0$ constraints also remain largely unchanged, whereas
for $\ell<1000$ they broaden, making these lower $H_0$ values compatible.
From Fig.~\ref{fig:boxplot}, there is a similar broadening in the running classes of
models r$\Lambda$CDM and rSB.   We return to the distinctions between these various
means of fitting TT residuals  at low $\ell$ in  Section \ref{sec:inflation} (see also Fig.~\ref{fig:residualslowl}).

\subsection{High $\ell\gtrsim 1000$ TT residuals}

In the full TT dataset, the lower $\Omega_c h^2$ and higher $H_0$ that is preferred by the low-$\ell$ residuals causes
even more significant in-phase high-$\ell$ deviations from the data.  Higher radiation driving beyond the 4th peak  implies
 sharper peaks than those seen in the data (see Fig.~\ref{fig:highLresid}).
Indeed, there remains in-phase oscillatory residuals of opposite sign at high $\ell>1000$ even with respect to the full $\Lambda$CDM TT maximum likelihood model.
The data favor smoother peaks than those that can be achieved in  the $\Lambda$CDM context,
producing the  so-called lensing anomaly, since additional lensing acts
to smooth the peaks (see Fig.~\ref{fig:response}).

 Specifically,  when allowing the amplitude of lensing
$C_\ell^{\phi\phi} \rightarrow A_L C_\ell^{\phi\phi}$ to float independently, $A_L= 1.22\pm 0.10$ \cite{Ade:2015xua}.  Alternately, if only $\ell>1000$ data are considered in the $\Lambda$CDM
context, $H_0=[64.1 \pm 1.7]$ \Hunits  as pointed out by Ref. \cite{addison2016}.  However,
given an inflationary power spectrum described by $A_s$ and $n_s$ as in  $\Lambda$CDM,
fitting the oscillatory residual in this way does not lead to a consistent global solution
\cite{aghanim2016}.

Relaxing the assumptions on the inflationary power spectrum allows for greater flexibility.
   In Fig.~\ref{fig:highLresid},  we see that moving to the $r\Lambda$CDM and $r$SB classes does indeed fit the oscillatory residuals better.   Correspondingly, in Fig.~\ref{fig:H0_lens},
these classes
allow even lower values of $H_0=66.6\pm 1.1$ and $66.2 \pm 1.2$ \Hunits respectively, corresponding to the higher values of
$\Omega_c h^2$, as shown in Fig.~\ref{fig:boxplot}.  Note also that this enhanced freedom
is also associated with a higher value of $\tau$ and, hence, $A_s$.  Given the suppressed curvature  power spectrum on large scales, it takes a larger $\tau$ to provide the same amplitude of polarization from reionization.

In Fig.~\ref{fig:ClPP}, we show that the combination of $A_s$ and $\Omega_c h^2$ changes in
the $r\Lambda$CDM model from Tab.~\ref{tab:modelstats}
 also enhances the lensing power spectrum.  However, lensing reconstruction data do
 not favor such an enhancement.     In Fig.~\ref{fig:H0_lens} we see that, correspondingly,  the preference for shifting $H_0$ even lower in the r$\Lambda$CDM model mainly goes away
 with
the TT$+\phi\phi$ dataset, where $H_0=67.2\pm 1.1$ \Hunits.

\begin{figure}
\centering
\includegraphics[width=0.4\textwidth]{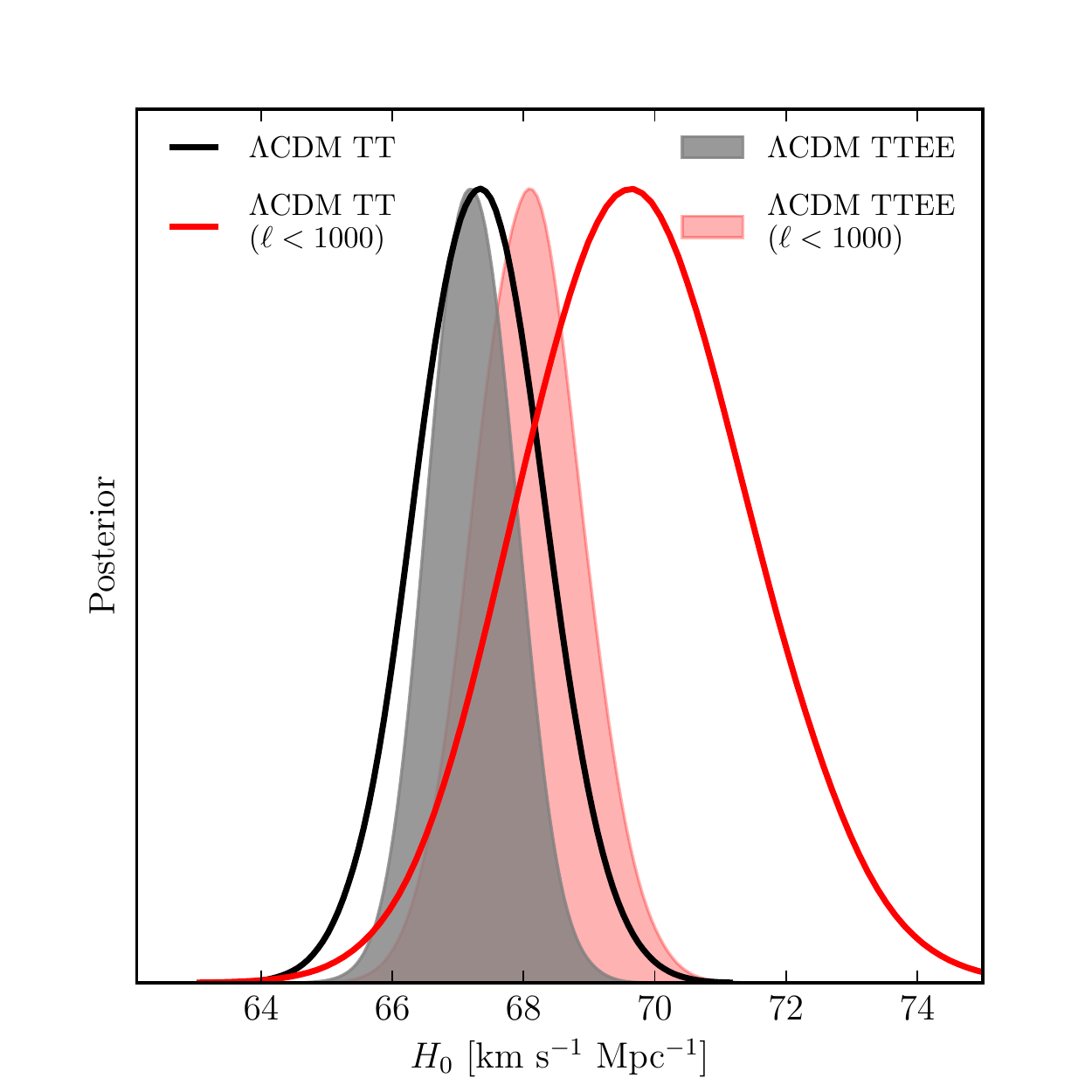}
\caption{\footnotesize $H_0$ and polarization data.   TE residuals
 shown in
Fig.~\ref{fig:TEresiduals}, especially near $\ell \sim 200$, drive $H_0$ lower, even in the
$\ell < 1000$ datasets (red curves to red shaded), whereas the full data range provide
values consistent with $\Lambda$CDM TT, with smaller errors. }
\label{fig:H0_pol}
\end{figure}

\begin{figure*}
\centering
\includegraphics[width=\textwidth]{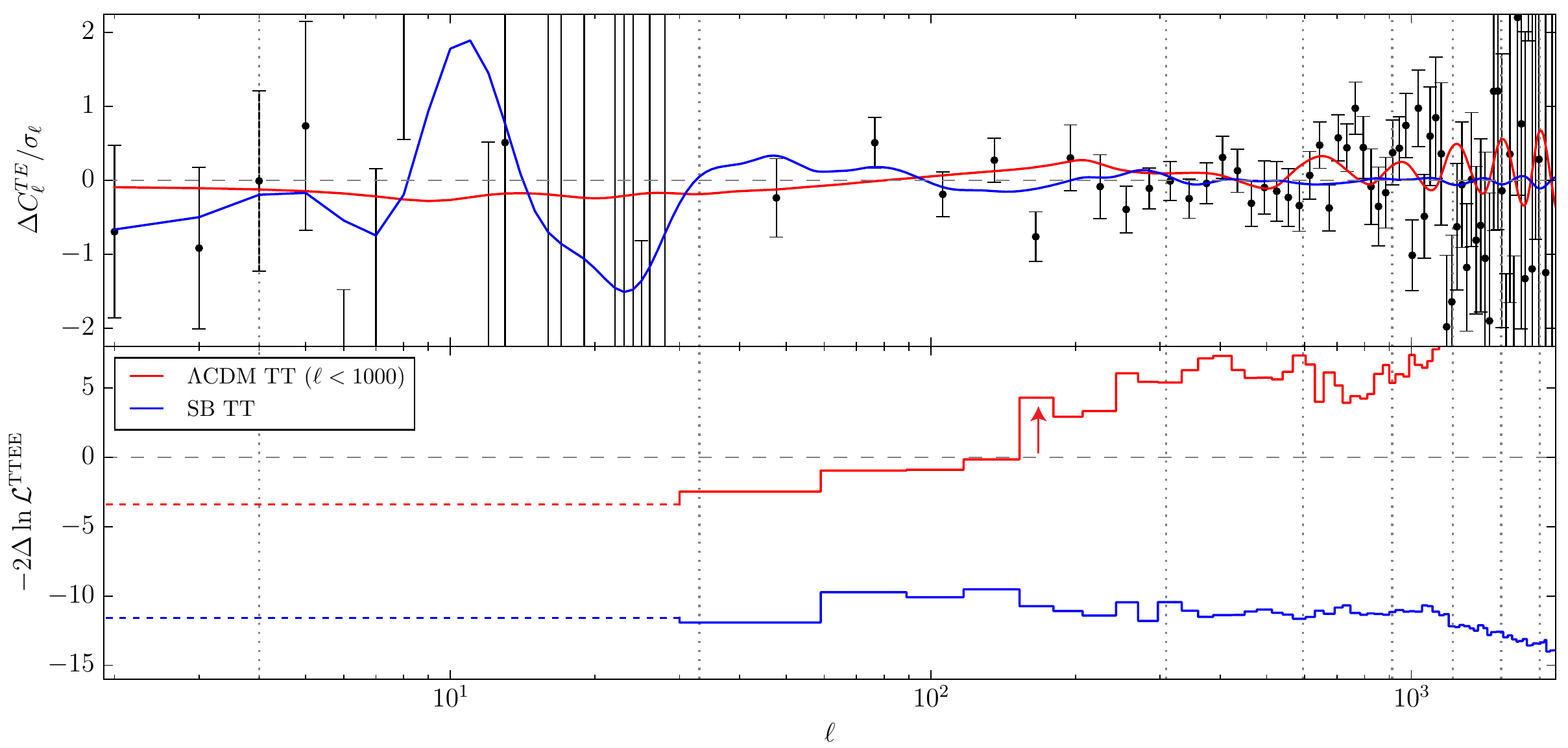}
\caption{\footnotesize Top panel: residuals in the TE data (points) and the same models as in Fig.~\ref{fig:TTresiduals} relative to the best fit $\Lambda$CDM TT model of Tab.~\ref{tab:modelstats} (dashed line).  Bottom panel: $2\Delta\ln {\cal L}^{\rm TTEE}$
 of models relative to
$\Lambda$CDM TT in the same form as Fig.~\ref{fig:TTresiduals}.
Cosmological parameters are fixed to their TT  best fit values and
the TT ($\ell<1000$) model also has the same foreground parameters as the TT model including polarized foregrounds.
The $\Lambda$CDM TT $(\ell < 1000)$ model becomes a worse fit to the TTEE dataset when passing
the outlier point at $\ell \approx 165$ (arrow),
whereas the SB TT model slightly improves the fit.
}
\label{fig:TEresiduals}
\end{figure*}

\subsection{Intermediate $\ell \sim 200$ TE residuals}

Ref.~\cite{Ade:2015xua} showed that the TE data alone constrain $H_0$ to comparable
precision as the full TT data, and that it favors the low $H_0$  of the latter as opposed
to the high $H_0$ of TT $(\ell < 1000)$.  We can
attribute some of this ability to the  enhanced, degeneracy breaking,
sensitivity to $\Omega_c h^2$ in the TE and EE spectra around the first polarization
trough at $\ell \sim 200$ (see
Fig.~\ref{fig:response}).    Given \Planck noise, TE is more constraining than EE, and so
we focus on its impact.

In Fig.~\ref{fig:H0_pol}, we show the  TTEE data set inferences on $H_0$ from the full
range and compare it to the $\ell<1000$ range.  Most of the extra information from polarization
comes from $\ell < 1000$, as expected given \Planck noise.  While discarding the $\ell>1000$ range in the TT analysis
doubles the $H_0$ errors, the impact on the TTEE analysis is notably smaller: {$H_0=68.1\pm 0.83$ \Hunits} compared with $67.2\pm 0.66$ \Hunits
 Moreover, the TE data at $\ell<1000$  pulls $H_0$ closer to the value preferred by
the  TT dataset of the full rather than $\ell <1000$ range.   On the face of it, this provides independent support for low $H_0$ that
does not involve the multipole ranges known to have anomalies in TT and provides some proof against instrumental systematics.
However, even given the large and sharp polarization responses to $\Omega_c h^2$ in Fig.~\ref{fig:response}, the  noisy TE \Planck data seem
unusually sensitive to $\Omega_c h^2$ and $H_0$.

To better understand the sensitivity, in Fig.~\ref{fig:TEresiduals}
we show the residuals of the TE data relative to the
best fit $\Lambda$CDM TT model and also compare them to the best fit $\Lambda$CDM TT ($\ell<1000$) model.   Note that neither cosmological model is  optimized to fit the polarization data.
Polarized and unpolarized foregrounds are jointly reoptimized in the $\Lambda$CDM
and SB TT cases, whereas for $\Lambda$CDM TT ($\ell <1000$) they are held fixed to their
reoptimized $\Lambda$CDM values.
The difference between the two models is relatively
small compared with the measured errors.   However, the data show a significant outlier in
the $\ell\sim 165$ bin relative to both models at $\sim 2\sigma$ for the full $\Lambda$CDM TT model (low $H_0$)
and $\sim 3\sigma$
for the $\Lambda$CDM TT ($\ell<1000$; high $H_0$) model.   We therefore attribute
the enhanced sensitivity to $H_0$ in large part to the presence of this one outlier.

To test this attribution,
we also show in Fig.~\ref{fig:TEresiduals} (bottom panel) the change in the TTEE likelihood between the $\Lambda$CDM TT and $\Lambda$CDM TT ($\ell <1000$) models
as a function of the maximum $\ell$.   The $\Lambda$CDM TT ($\ell<1000$) model jumps from a better fit to a worse fit across the  $\ell=165$ bin, unlike with the TT likelihood (cf.~Fig.~\ref{fig:TTresiduals}).
We also show the best fit SB TT model in Fig.~\ref{fig:TEresiduals}.   Intriguingly, even though this model is not optimized to fit the outlier, it does marginally reduce tension with the $\ell=165$ bin.
It is also interesting to note that even in the r$\Lambda$CDM
where $H_0=66.9\pm 0.69$ \Hunits and rSB classes where $H_0=66.8\pm 0.72$ \Hunits, the TTEE
data set no longer allows the even lower $H_0$ values found from fitting the $\ell>1000$ oscillatory
TT residuals even with no lensing reconstruction data (see Fig.~\ref{fig:boxplot}).

However,  the \Planck collaboration considers the TE and EE data sets as
preliminary and so the anomalous sensitivity to $H_0$ in the $\Lambda$CDM context,
 while intriguing, should not be overinterpreted.

\begin{figure}
\centering
\includegraphics[width=0.4\textwidth]{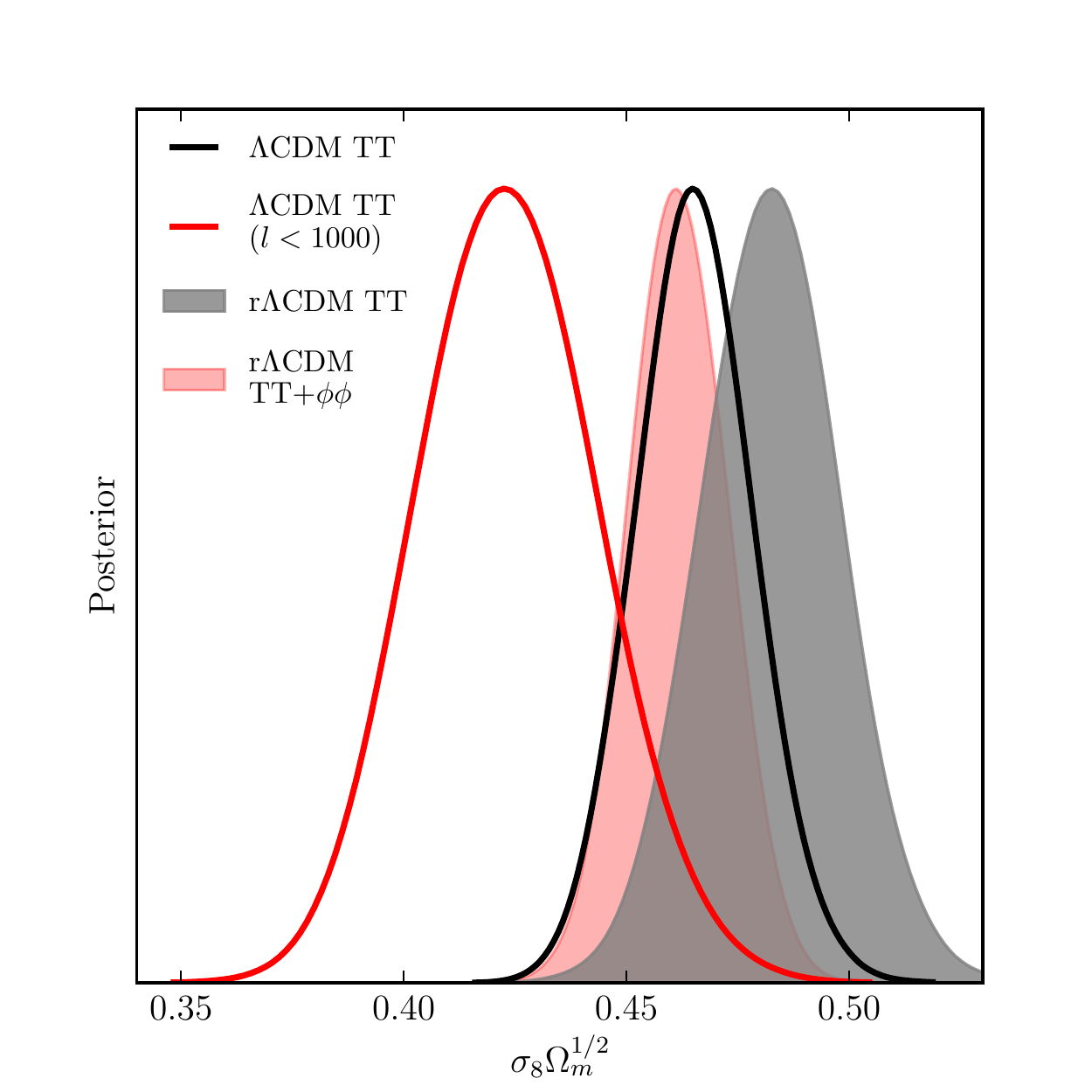}
\caption{\footnotesize Amplitude of structure $\sigma_8 \Omega_m^{1/2}$.   Shifts in $\Omega_c h^2$
from $\Lambda$CDM TT $(\ell <1000$; red curve) to the full range (black curve) drive
$\sigma_8 \Omega_m^{1/2}$ higher.  Fitting the in-phase high-$\ell$ residuals with the
enhanced freedom of the r$\Lambda$CDM model drives it even higher (gray shaded), but in a
manner inconsistent with the lensing reconstruction data (red shaded). }
\label{fig:sigma8}
\end{figure}

\subsection{Local amplitude $\sigma_8\Omega_m^{1/2}$}

The lowering of $H_0$ and raising of $\Omega_m$ between the WMAP9 or \Planck TT $(\ell < 1000)$ cosmology
and the \Planck (full) TT cosmology causes an increase in the amount of local structure
$\sigma_8 \Omega_m^{1/2}$ (see Fig.~\ref{fig:sigma8}).  This increase is in moderate tension with measurements of the local cluster abundance, depending on the mass calibration employed
\cite{Ade:2013lmv,Ade:2015xua}, and of weak gravitational lensing \cite{Joudaki:2016mvz,Hildebrandt:2016iqg}.

In Fig.~\ref{fig:sigma8}, we show the effect of generalizing the model class to r$\Lambda$CDM.
The allowed further raising of $\Omega_c h^2$ and lowering of $H_0$ with just the TT data set
 shifts
$\sigma_8\Omega_m^{1/2}$ to higher values.  However, once the $\phi\phi$ lensing reconstruction data are included, this preference disappears.

\begin{figure*}
\centering
\includegraphics[width=0.99\textwidth]{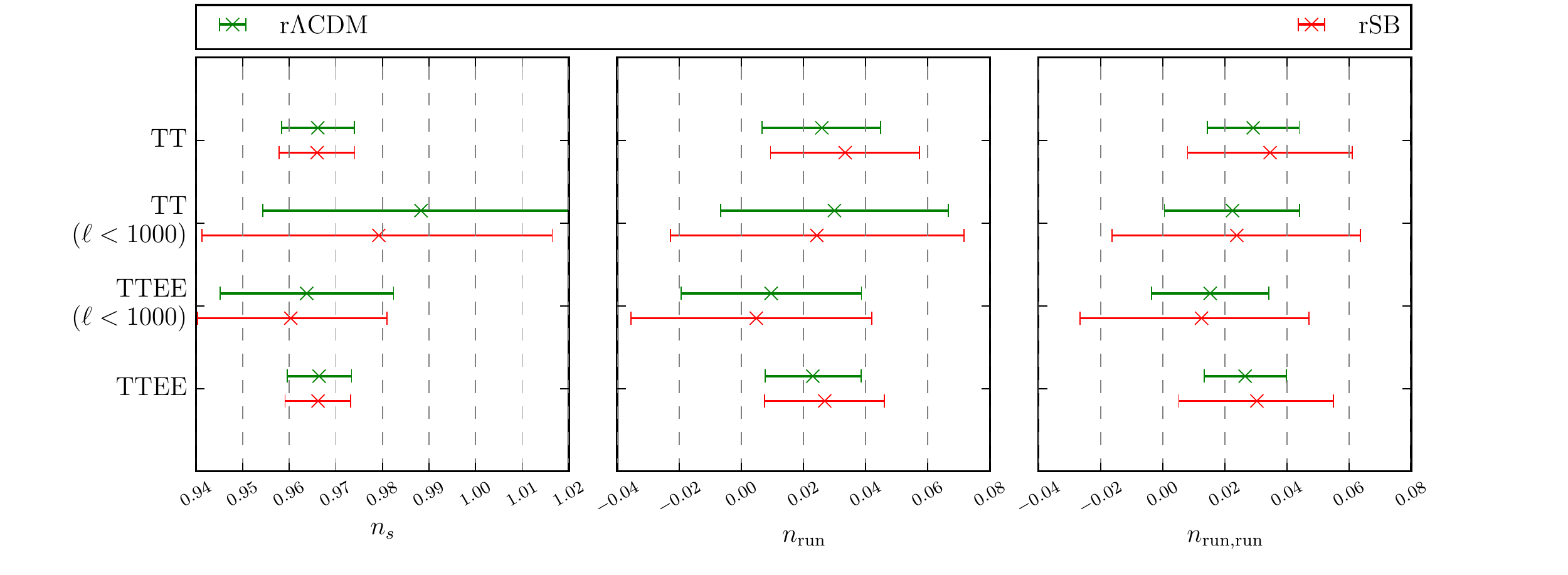}
\caption{\footnotesize Constraints on inflationary parameters -- tilt, its running, and the running of the running from Eq.~\ref{eqn:running} -- for the 4 combinations of datasets
in Tab.~\ref{tab:data}.}
\label{fig:boxplotrunning}
\end{figure*}

\begin{figure}
\centering
\includegraphics[width=0.49\textwidth]{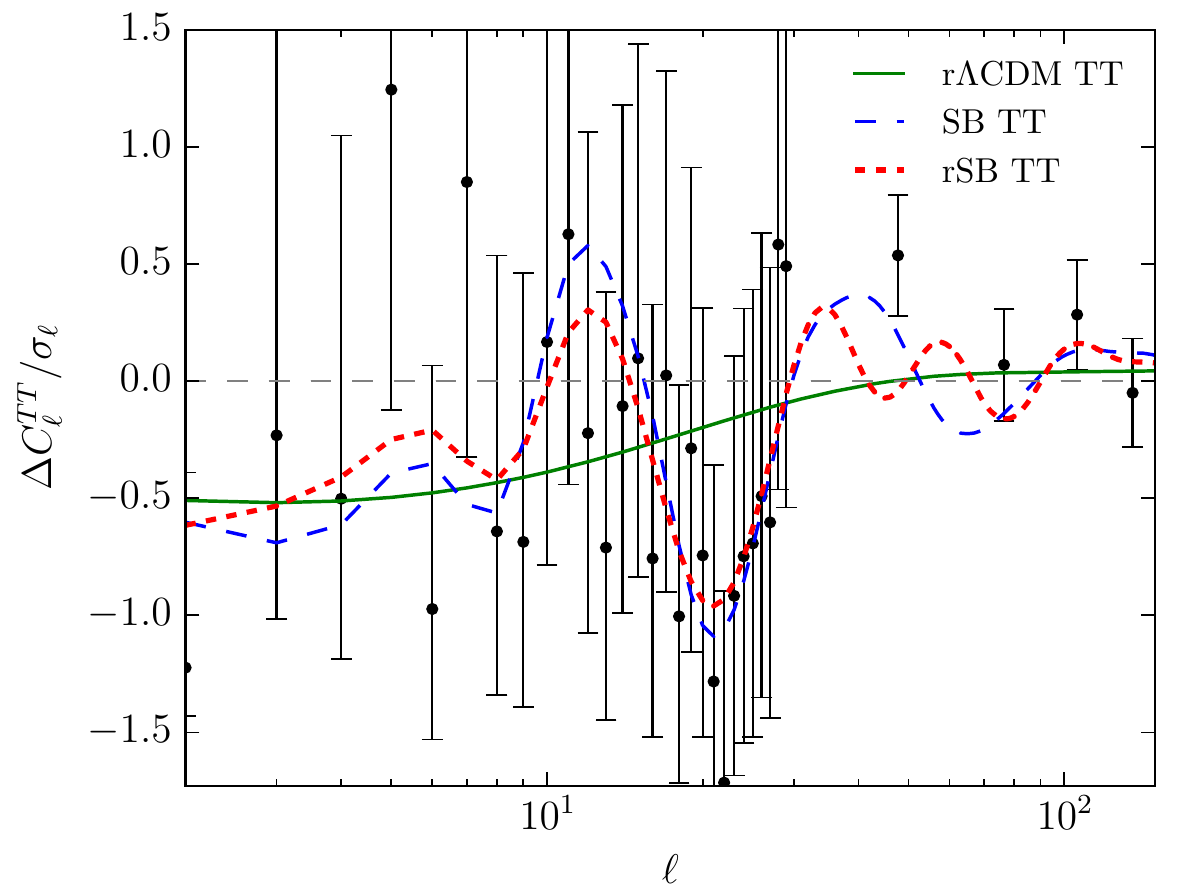}
\caption{\footnotesize Low-$\ell$ TT residuals  (points) and models relative to the best fit $\Lambda$CDM TT model of Tab.~\ref{tab:modelstats} (black dashed).  While the running model r$\Lambda$CDM TT (green solid) can suppress power, the SB TT model (blue dashed) accommodates the sharp dip favored by the data.   Allowing both
running and SB in the rSB TT model (red dotted) leads to a similar fit as SB TT. }
\label{fig:residualslowl}
\end{figure}

\begin{table}
  \begin{tabular}{|c|c|c|c|c|}
    \hline
    \input{table2_3.tex}
  \end{tabular}
  \caption{\footnotesize Inflationary parameters of best fit models for various datasets. The running parameters are defined in Eq.~\ref{eqn:running} and  $m_1$, $m_2$, $m_3$ are are the first 3 principal components of the SB parameters rank ordered by increasing variance. }
   \label{tab:modelstats2}
\end{table}

\begin{figure}
\centering
\includegraphics[width=0.49\textwidth]{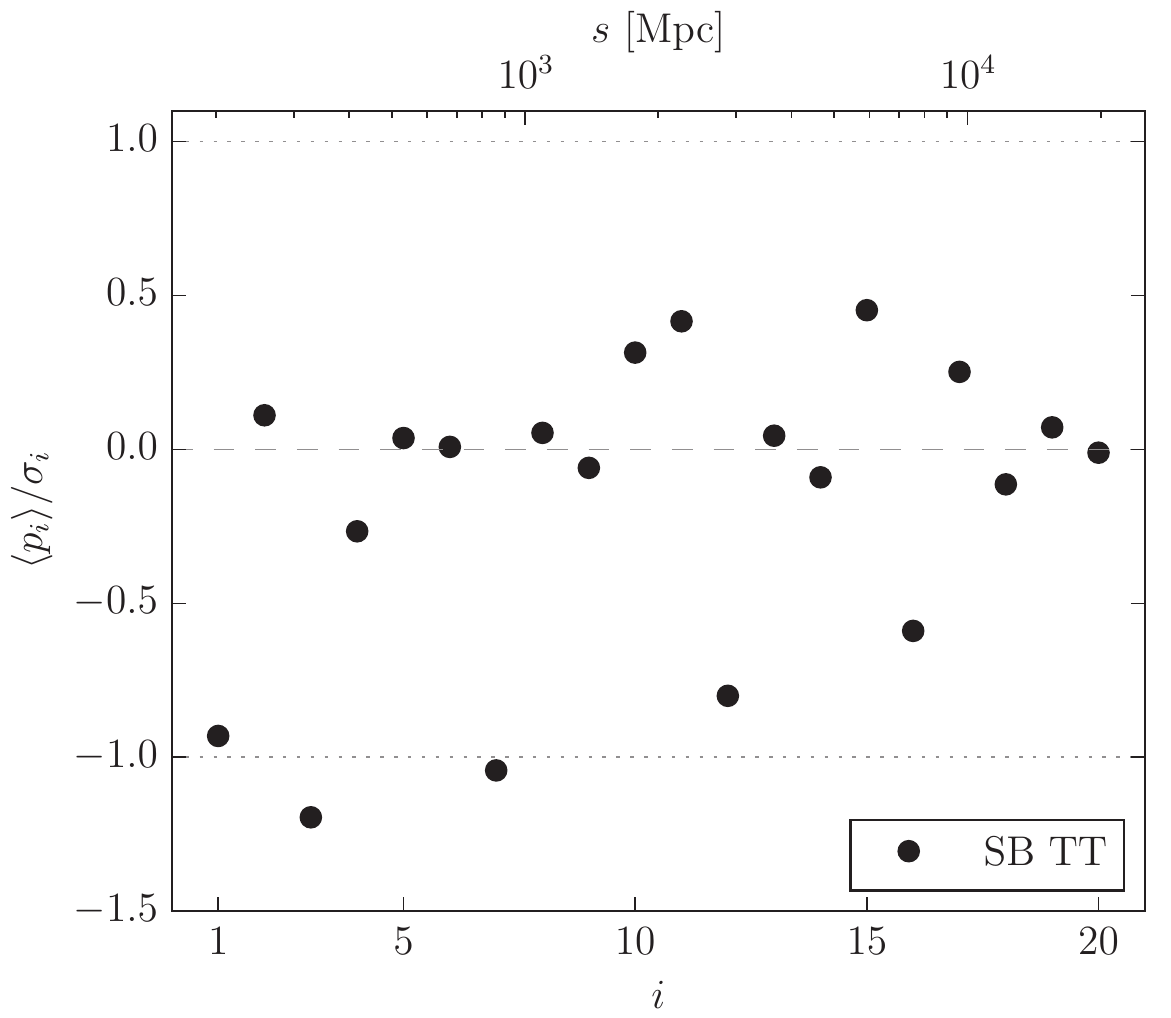}
\caption{\footnotesize SB parameters of the SB TT analysis in units of their errors.   No single SB parameter $p_i$ is significantly detected due to covariances hidden in the marginalized errors.   }
\label{fig:SBparameters}
\end{figure}

\begin{figure}
\centering
\includegraphics[width=0.48\textwidth]{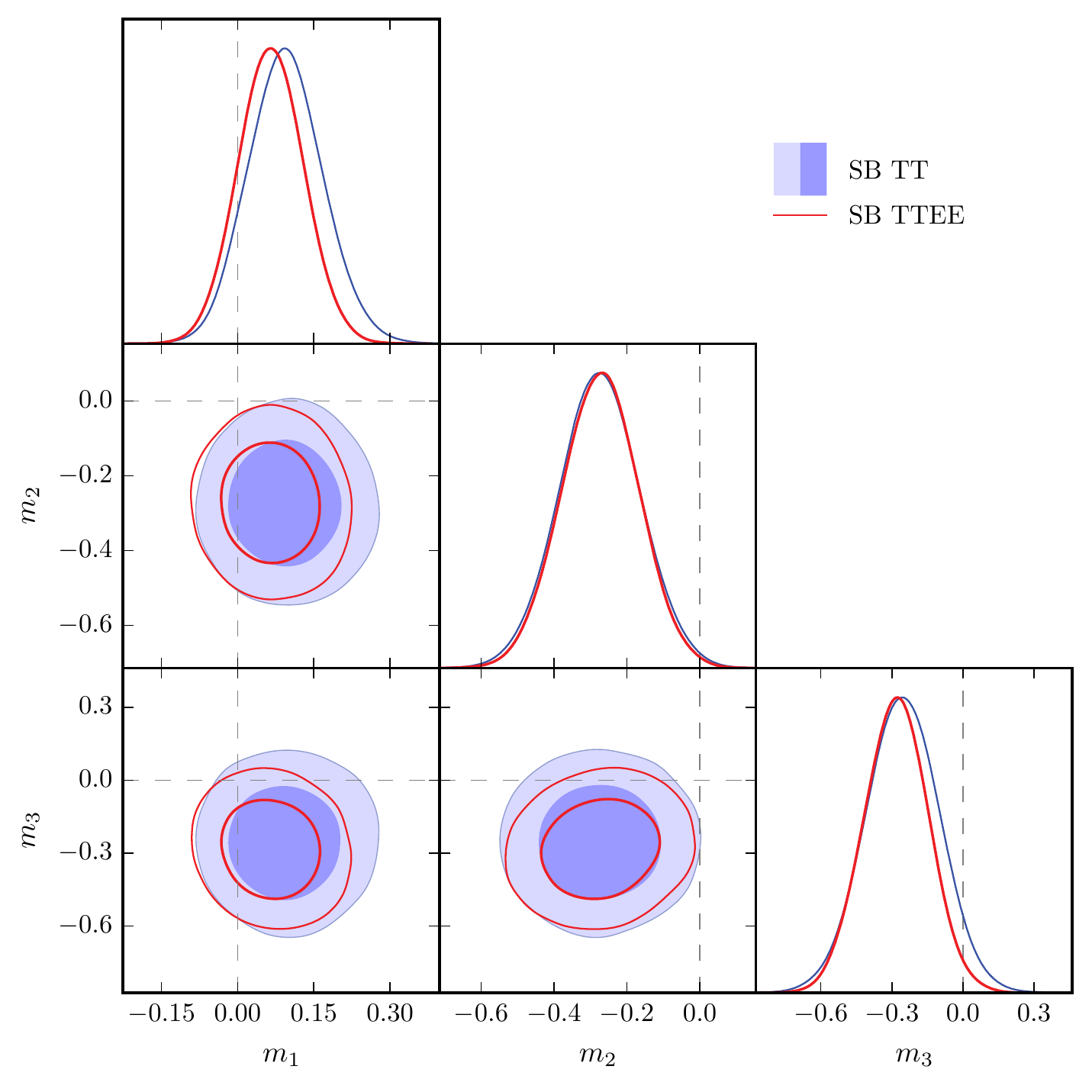}
\caption{\footnotesize First 3 principal components of the SB TT $p_i$ covariance matrix in the TT (thin blue) and TTEE (thick red) analysis (68\%, 95\% CL).   Both $m_2$ and $m_3$ show non-zero deviations whose
significance is slightly enhanced by the addition of polarization information.}
\label{fig:PCtriangle}
\end{figure}

 \begin{figure}
\centering
\includegraphics[width=0.44\textwidth]{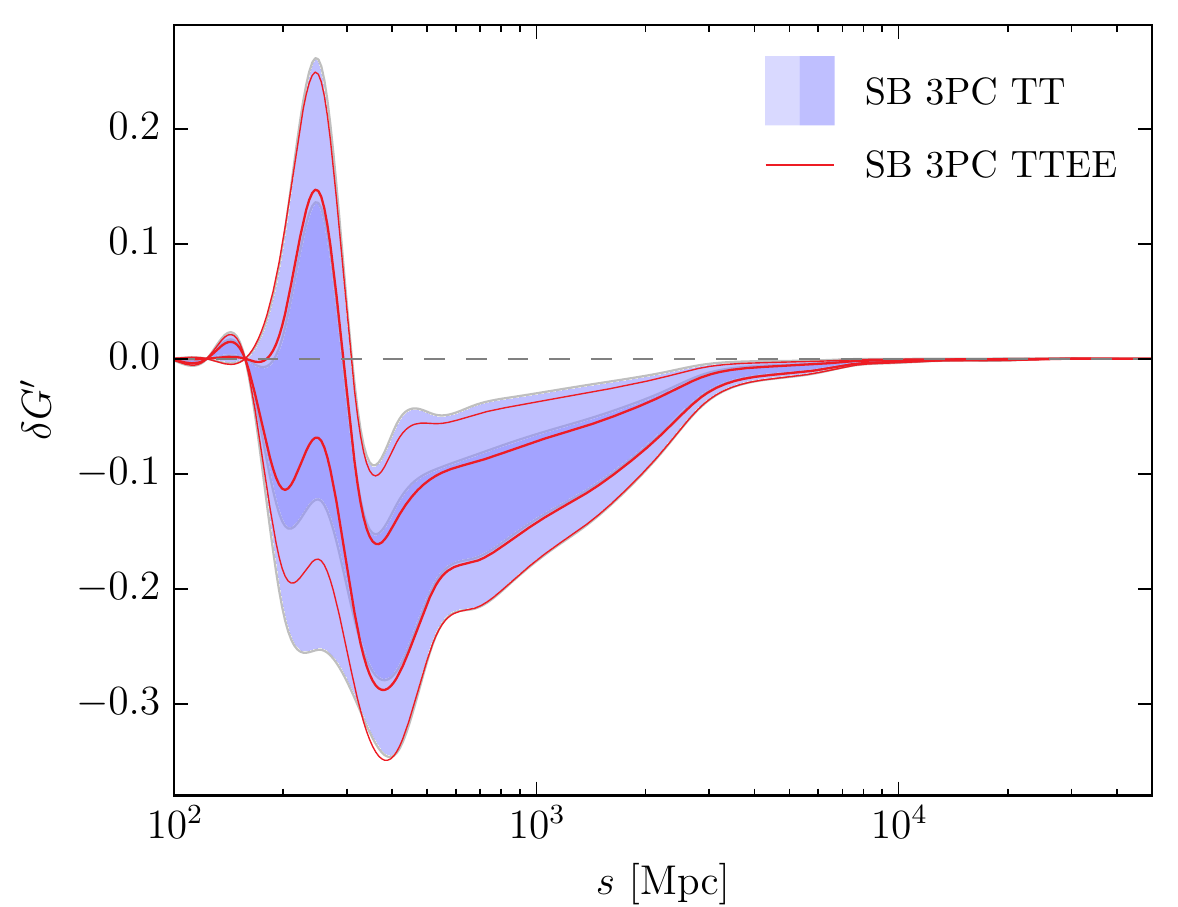}
\caption{\footnotesize Functional reconstruction of generalized tilt  ($\delta G'$: 68\%, 95\% CL)
 from the first 3 principal components
in the SB TT (shaded) and SB TTEE (dashed and dotted lines) analyses.   Polarization information
slightly sharpens the dip near 400\,Mpc.
}
\label{fig:dGp_3PCs}
\end{figure}

\begin{figure}
\centering
\includegraphics[width=0.42\textwidth]{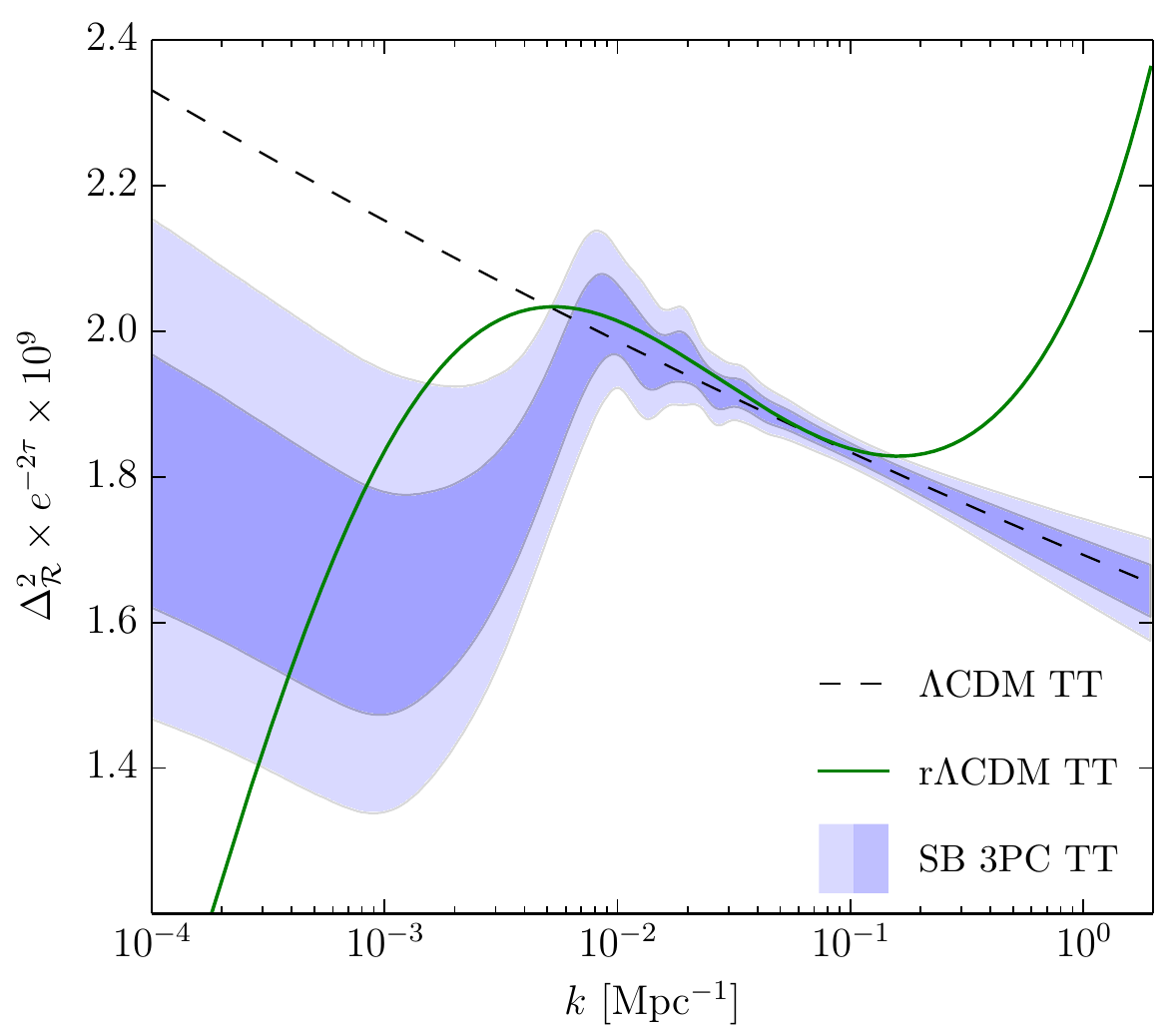}
\caption{\footnotesize Functional reconstruction of inflationary power spectrum
 ($\Delta_\curv^2$: 68\%, 95\% CL)
 from the first 3 principal components
in the SB TT (shaded) analysis compared with the best fit running model r$\Lambda$CDM TT.}
\label{fig:Delta2PC}
\end{figure}

\section{Inflationary Features}
\label{sec:inflation}

 The same residuals in the \Planck TT data that drive shifts in $H_0$ and $\sigma_8 \Omega_m^{1/2}$ provide hints that inflation may be more complex than single-field slow-roll inflation would imply.
 No single set of $\Lambda$CDM parameters with a power-law curvature power spectrum
   simultaneously gives a very good fit to the
 TT data at  low $\ell$ and high $\ell$, as shown in Fig.~\ref{fig:TTresiduals}.

 Although single-field inflation allows for running of the tilt and running of the running, the slow-roll approximation predicts
 that each order should be successively suppressed by ${\cal O}(n_s-1)$, i.e.\ $n_{\rm run,run}
 \lesssim 10^{-4}$.
 However, in the r$\Lambda$CDM analysis $n_{\rm run,run}= 0.029 \pm 0.015$
 \cite{Ade:2015xua}, a $1.9\sigma$ preference for a violation of slow-roll inflation (see Fig.~\ref{fig:boxplotrunning}).  In Tab.~\ref{tab:modelstats2}, we give the best fit values
 for running parameters in the various model classes, and in Fig.~\ref{fig:residualslowl} we show
 how well these models fit the data residuals at low $\ell$.   While the r$\Lambda$CDM model
 does indeed lower power on large scales, it fails to fit the sharp decline in power at
 $\ell \sim 20-30$.  As shown in Fig.~\ref{fig:highLresid}, there is also a slight preference
 for nonzero running parameters due to their ability to accommodate cosmological parameter changes that fit the high-$\ell$ oscillatory residuals.   However, this preference does not remain
 if the lens reconstruction data are considered.

 Given the large preferred values of $n_{\rm run,run}$, such models violate slow roll and may in their own right be considered inflationary features, so there is no
 theoretical motivation to model the inflationary curvature power spectrum as a truncated
 Taylor expansion
 around the pivot point as in Eq.~(\ref{eqn:running}).    Indeed, we see
 in Fig.~\ref{fig:residualslowl} that the SB parameters of the generalized tilt $\delta G'$ allow a sharper decline at  $\ell \sim 20-30$, which better fits the data.  On the other hand, with 20 free parameters, the SB model tends to
 overfit the residuals  leaving highly correlated parameter errors and no single $p_i$ with more than a $\sim 1\sigma$ deviation
 from the slow roll prediction of zero (see Fig.~\ref{fig:SBparameters}).

 From Fig.~\ref{fig:TTresiduals}, we see that the net improvement in fitting the residuals with SB at $\ell < 30$
 is $2\Delta\ln {\cal L}^{\rm TT} \sim 12$.  We also show that combining both SB and running parameters leads to fits that are very similar to SB at low $\ell$.
  Correspondingly, once the SB parameters are marginalized, the significance of the preference for finite values decreases to  $n_{\rm run,run}=0.035\pm 0.027$  (see Fig.~\ref{fig:boxplotrunning}).  Its best fit value $0.041$ actually increases in rSB, in part due to its ability to fit the
  high-$\ell$ in-phase residuals.

 In order to expose the most significant linear combination of the SB $p_i$ parameters and their implications,
 we examine the $m_a$ principal components (PCs) of its covariance matrix $C_{ij} = \langle p_i p_j \rangle$, rank ordered by increasing eigenvalue or variance, following Ref. \cite{miranda2015}, as detailed in the Appendix.   For the SB TT analysis, the best constrained  PCs also
 contain the most significant deviations from slow roll.   Note that $m_2=m_3=0$ is outside
 the $95\%$ CL region and is in fact excluded at $\sim 99\%$ CL in Fig.~\ref{fig:PCtriangle} with {$m_2=-0.27\pm 0.11$ and $m_3=-0.25\pm 0.16$}.  Of course one should interpret this as a local significance given the number of well measured PCs that this anomaly could have appeared in.
 Furthermore, using the SB TT PC basis,
 we can assess whether the additional polarization information in the TTEE data set supports
 or weakens this preference.   In Fig.~\ref{fig:PCtriangle} we see that polarization
 enhances the significance of the deviation to $2\sigma$ in  $m_3={-0.28\pm 0.14}$
 while leaving $m_2={-0.27\pm 0.11}$ essentially unchanged at $\sim 2.5\sigma$  and improving the bounds on $m_1$.

 Retaining (or ``filtering") the SB to only the first 3 PCs (``SB 3PC TT", see Eq.~\ref{eqn:PCfilter}), we can better isolate the most significant
 implications for inflation and the curvature power spectrum.   In Fig.~\ref{fig:dGp_3PCs},
 we show the inflationary implications for $\delta G'$, the analog of the tilt for models with
 rapid deviations from slow roll.   The 3PC preference is for a $>95\%$ CL sharp suppression in $\delta G'$
 beginning at $\sim 300$ Mpc and preceded by a less significant enhancement.   While the additional polarization information in TTEE does not
 further enhance the amplitude or significance of the suppression, it does enhance the sharpness.

 Finally, in Fig.~\ref{fig:Delta2PC} we show the implications for the curvature power spectrum.
 Unlike the smooth deviations of the  r$\Lambda$CDM TT model, the SB 3PC TT model has
 a sharp suppression around $k \approx 0.004$ Mpc$^{-1}$.  In canonical single-field inflation, such a suppression would be associated with an abrupt transition from
 a faster to a slower rolling rate.

\section{Discussion}
\label{sec:discussion}

We have explored the interplay between  features from inflation, features in the \Planck 2015 data, and shifts in the cosmological
parameters.

Preference for high $H_0$ values in $\Lambda$CDM from the TT data at $\ell<1000$ is driven by low $\ell<40$ residuals in the data, which then prefer the acoustic peaks to be raised by enhancing
radiation driving through lowering $\Omega_c h^2$, and consequently raising $H_0$
to match the angular position of the peaks.   However, these cosmological parameter
variations do not match the residuals particularly well.   If we marginalize their impact on
$H_0$ by fitting, instead, to inflationary features of the generalized tilt $\delta G'$ form in
the SB analysis, then the low $H_0$ favored by the full TT data is compatible with
even the  $\ell <1000$ TT data.

Correspondingly, the SB parameters show 2 principal
components that jointly deviate from zero at the $\sim 99\%$ CL in local significance.
These deviations in $\delta G'$ prefer a sharp suppression of power around $k\approx 0.004$ Mpc$^{-1}$.   The residuals can also be fit to a running of the running of the tilt in
the r$\Lambda$CDM model, but those fits cannot reproduce the sharpness of the
suppression.  Furthermore, the large running of the running favored would itself represent
an inflationary feature in the power spectrum that cannot be explained in slow-roll inflation.

The same $\ell>1000$ oscillatory TT residuals which drive $H_0$ lower and $\sigma_8 \Omega_m^{1/2}$ higher in the full TT data set prefer an even lower $H_0$ if only
 $\ell >1000$ is considered \cite{addison2016}, but this
 does not provide a good  global fit in $\Lambda$CDM with power-law initial conditions
 \cite{aghanim2016}.  With running parameters in the tilt, $H_0=66.6\pm 1.1$ \Hunits.   The combination of low-$\ell$ and high-$\ell$ anomalies makes running of the running preferred at the $1.9\sigma$ level in the TT data set.
However, on the low $\ell$ side, once the sharp feature is marginalized with SB this preference drops to the $1.3\sigma$  level.
On the high $\ell$ side, once lensing reconstruction is taken into account, preference for fitting oscillatory residuals disappears and $H_0 = 67.9\pm 0.92$ \Hunits returns to nearly the $\Lambda$CDM value.

In polarization, the signature of changes in $H_0$ through $\Omega_c h^2$ is much sharper
and somewhat larger.    Polarization suffers less projection effects and lacks the early ISW that broadens the signature in temperature and its measurement at $\ell \sim 200$ is important for assuring the robustness of $H_0$ inferences from the CMB in $\Lambda$CDM.

Indeed, despite being noisier than TT,  \Planck TE data alone already constrain $H_0$ as
well as  TT \cite{Ade:2015xua}.
They also are consistent with the low $H_0$ constraints from the full TT data,
even though their impact comes from $\ell<1000$.
On the other hand, the \Planck TE data are anomalously sensitive to $H_0$ due to an outlier at $\ell \sim 165$, which even more strongly disfavors the high $H_0$ of the $\ell<1000$ TT data fit.

Polarization data are also beginning to help constraints on the inflationary features favored by the low-$\ell$ TT residuals.   In \Planck 2015, only the $\ell\ge 30$ HFI data was released, and so the main effect is to strengthen the confinement of the features to large scales, thus favoring a sharper feature in the joint analysis.  This sharpening in turn enhances the significance
of the deviations in one of the principal components to $2\sigma$ leaving the other at $2.5\sigma$.
At $\ell <30$, only LFI polarization data  was released, and the impact of inflationary features
is to change inferences on the optical depth $\tau$, since with smaller inflationary power a higher
$\tau$ is required to produce the same polarization power.   In the final \Planck release, the inclusion of HFI polarization data at $\ell <30$ should also provide insight on the low-$\ell$ TT residuals and their implications for inflationary features.   It will be important then to distinguish
inflationary features from reionization features beyond the simple near instantaneous reionization
models used here \cite{Heinrich:2016ojb}. We leave these considerations to a future work.

\acknowledgments
WH and CH were supported by grants
NSF PHY-0114422,
NSF PHY-0551142,
U.S.~Dept.\ of Energy contract DE-FG02-13ER41958, and
NASA ATP NNX15AK22G. Computing resources were provided by the University of Chicago Research Computing Center. VM was supported in part by the Charles E.~Kaufman Foundation, a supporting organization of the Pittsburgh Foundation.

\appendix
\label{sec:appendix_GSR}

\section{GSR Spline Basis}
\label{app:GSR}

The GSR approach \cite{Stewart:2001cd} is a generalization of the slow-roll approximation that allows for
features in the power spectrum of $\Delta \ln k \lesssim 1$ in a general single-field model
that includes canonical \cite{Dvorkin:2009ne}, $P(X,\phi)$ \cite{Hu:2011vr},
and Horndeski (Galileon)  inflation \cite{Motohashi:2017gqb}.
In this approximation, the curvature power spectrum $\Delta_\curv^2$ is sourced by a single
function $f=2\pi z \sqrt{c_s} s$, where $z$ is the Mukhanov variable in the Mukhanov-Sasaki equation, $c_s$ is the sound speed and $s$ is the sound horizon:
\begin{align}\label{eqn:def_sound_horizon}
	s = \int_{a}^{a_{\rm end}} \frac{da}{a} {c_s \over aH}.
\end{align}
Here $a_{\rm end}$ is the scale factor at the end of inflation and $H$ is the Hubble parameter during inflation. For a canonical scalar field in Planck units $\phi$, $z= a^2 d\phi/d a$.

In the ordinary slow-roll approximation,  $\Delta_\curv^2 \approx f^{-2}$, whereas in GSR:
\begin{align} \label{eqn:GSRpower}
\ln \Delta_\curv^{2}(k) &\approx  G(\ln s_{*}) + \int_{s_{*}}^\infty {d s\over s} W(ks) G'(\ln s)\\
&\quad + \ln \left[ 1+ I_1^2(k) \right], \nonumber
\end{align}
where
\begin{equation}
G (\ln s)= - 2 \ln f    + {2 \over 3} (\ln f )'   ,
\end{equation}
with $' = d/d\ln s$ and  $s_*$ is an arbitrary epoch during inflation such that all relevant $k$-modes
are well outside the sound horizon, $k s_* \ll 1$.  The window functions,
\begin{align}
	\label{eqn:powerwindow}
W(x) &= {3 \sin(2 x) \over 2 x^3} - {3 \cos (2 x) \over x^2} - {3 \sin(2 x)\over 2 x},
\end{align}
at leading order  and
\begin{align}
X(x) & = {3 \over x^3} (\sin x - x \cos x)^2 ,
\end{align}
at second order through
\begin{eqnarray}
I_1(k) &=& { 1\over \sqrt{2} } \int_0^\infty {d s \over s} G'(\ln s) X(ks),
\end{eqnarray}
 characterize the freezeout of the source function around sound horizon crossing.
   This form for the power spectrum remains a good approximation
if the second order term \cite{Dvorkin:2011ui}
\begin{equation}
I_1 < \frac{1}{\sqrt{2}} ,
\label{eq:I1prior}
\end{equation}
and hence allows for up to order unity features in the curvature power spectrum.

The ordinary slow-roll approximation corresponds to $G'(\ln s) = 1-n_s$, and results in
a power-law curvature power spectrum.
In the main text, we use GSR to fit the low-$\ell$ anomalies in the power spectrum,
and so we choose to restrict our parameterization of fluctuations,
$\delta G'$, around this constant  to
\begin{equation}
200 < \frac{s}{\rm Mpc}< 20000.
\label{eqn:range}
\end{equation}
Next, we follow Ref. \cite{Dvorkin:2011ui} in defining a band limit for the frequency of
deviations by sampling $p_i = \delta G'(\ln s_i)$ at a rate of 10 per decade for a total of 20
parameters.  This rate is sufficient to capture the low-$\ell$ anomalies.
We then construct the smooth function using the natural spline of these points.

Specifically, we exploit the linearity of splines to define the spline basis (SB) functions $B_i(\ln s)$ from the natural spline of the
set of unit amplitude perturbations $p_i=1$, $p_{j\ne i}=0$.
In the SB class of models we then describe the source function as:
\begin{align}
 G'(\ln s)  = (1-n_s)+  \sum_i p_i B_i(\ln s).
\label{eqn:Gpbasis}
\end{align}
The advantage of the GSR form in Eq.~(\ref{eqn:GSRpower}) is that the integrals
are linear in $G'$ and, hence, the impact of the individual components can
be precomputed separately as:
\begin{align}
W_i(k) &= \int_{s_*}^{\infty} {d s\over s} W(k s) B_i(\ln s) , \nonumber\\
X_i(k) &= \int_{0}^{\infty}  {d s\over s} X(k s) B_i(\ln s),
 \label{eqn:gsr_window_integrals}
\end{align}
so that the power spectrum becomes a sum over the basis
\begin{align} \label{eqn:ps2v2_basis}
	\ln \Delta_{\mathcal{R}}^2(k) =&  \ln A_s\left(\frac{k}{k_0}\right)^{n_s-1} 	+ \sum_i p_i \big[ W_i(k) - W_i(k_0) \big] \nonumber\\
	&+ \ln \left[ \frac{1+ I_1^2(k)}{1+ I_1^2(k_0)} \right],
\end{align}
where
\begin{align}
I_1(k) =  \frac{ \pi}{2\sqrt{2}}(1-n_s) +\frac{1}{\sqrt{2}} \sum_i p_i X_i(k).
\end{align}
Note that we have absorbed the normalization constant $G(\ln s_*)$ into the
amplitude of the power spectrum at the scale $k_0$:
\begin{align}
A_s =  \Delta_{\curv}^2(k_0).
\end{align}
In the rSB class of models we replace the first term in Eq.~(\ref{eqn:ps2v2_basis})
with the running form defined by Eq.~(\ref{eqn:running}).

Since our choice of  parameters oversamples $\delta G'$ relative to what the data can constrain,  individual measurements of $p_i$ from the MCMC mainly fit noise with any true signal
buried in the covariance between parameters.    We therefore  construct the principal
components derived from an eigenvalue decomposition of the MCMC covariance matrix estimate:
\begin{align}
C_{ij} &= \langle p_i p_j \rangle - \langle p_i \rangle \langle p_j \rangle \nonumber\\
 &= \sum_a S_{ia} \sigma_a^2 S_{ja},
\end{align}
where $S_{ia}$ is an orthonormal matrix of  eigenvectors.   Specifically, we define
the PC parameters
\begin{align}
m_a = \sum_i S_{ia} p_i,
\end{align}
such that their covariance matrix satisfies:
\begin{align}
\langle m_a m_b \rangle - \langle m_a \rangle \langle m_b \rangle = \delta_{a b} \sigma_a^2.
\end{align}
Given a rank ordering of the PC modes from smallest to largest variance, we can
also construct a 3 PC filtered reconstruction of $\delta G'$ as in Ref. \cite{Dvorkin:2011ui}:
\begin{equation}
\delta G'_{3\rm PC}(\ln s_i) = \sum_{a=1}^{3} m_a S_{ia},
\label{eqn:PCfilter}
\end{equation}
and similarly use $\delta G'_{3 \rm PC}$ to construct the 3PC filtered $\Delta^2_\curv$ used in
the main text.

\bibliography{InfPC}

\end{document}

%% file: table1_3.tex
Parameter & $\Lambda$CDM TT & $\Lambda$CDM TT $(\ell < 1000)$ & r$\Lambda$CDM TT & r$\Lambda$CDM  TT\,+\,$\phi\phi$ & SB TT & rSB TT \\
\hline \hline
$100\Omega_b h^2$ & 2.22 & 2.26 & 2.20 & 2.20 & 2.23 & 2.20 \\
$\Omega_c h^2$ & 0.120 & 0.115 & 0.121 & 0.120 & 0.121 & 0.121 \\
$100\theta_{MC}$ & 1.0409 & 1.0408 & 1.0407 & 1.0409 & 1.0409 & 1.0407 \\
$100\tau$ & 7.61 & 7.51 & 8.66 & 6.39 & 9.02 & 7.96 \\
${\rm{ln}}(10^{10} A_s)$ & 3.07 & 3.06 & 3.09 & 3.04 & 3.10 & 3.08 \\
$n_s$ & 0.965 & 0.980 & 0.970 & 0.972 & 0.962 & 0.970 \\
\hline
$H_0$ [km s$^{-1}$ Mpc$^{-1}$]& 67.3 & 69.4 & 66.5 & 67.1 & 67.0 & 66.5 \\
$\sigma_8 \Omega_m^{1/2}$ & 0.465 & 0.432 & 0.483 & 0.463 & 0.476 & 0.480 \\
\hline

%% file: table2_3.tex
Parameter & r$\Lambda$CDM TT & r$\Lambda$CDM TT + $\phi\phi$ & SB TT & rSB TT \\
\hline \hline
$n_{\rm run}$ 		& 0.033 & 0.030 & $ - $ & 0.042 \\
$n_{\rm run, run}$ 	& 0.032 & 0.026 & $ - $ & 0.041 \\
$m_1$ 				& $ - $ & $ - $ & 0.078 	& 0.025 \\
$m_2$ 				& $ - $ & $ - $ & -0.34 	& -0.085 \\
$m_3$ 				& $ - $ & $ - $ & -0.35 	& -0.12 \\
\hline